\documentclass{ws-ijmpd}

\def\ltsim{\hbox{\raise 2pt \hbox {$<$} \kern-1.1em \lower 4pt \hbox {$\sim$}}}
\def\ltapprox{\hbox{\raise 2pt \hbox {$<$} \kern-1.1em \lower 5pt \hbox 
{$\approx$}}}
\def\gtsim{\hbox{\raise 2pt \hbox {$>$} \kern-1.1em \lower 4pt \hbox {$\sim$}}}
\def\gtapprox{\hbox{\raise 2pt \hbox {$>$} \kern-1.1em \lower 5pt \hbox 
{$\approx$}}}

\def\arcsec{$^{\prime\prime}$}
\def\arcmin{$^{\prime}$}
\def\degrees{$^{\circ}$}
\def\srm{$\sigma_{\rm RM}$}
\def\rmm{$\langle{\rm RM}\rangle$}
\def\absrmm{$\arrowvert \langle {\rm RM} \rangle  \arrowvert$}

\begin{document}
\markboth{F. Govoni, L. Feretti}
{Magnetic fields in clusters of galaxies}

%
\catchline{}{}{}{}{}
%

\title{MAGNETIC FIELDS IN CLUSTERS OF GALAXIES}

\author{FEDERICA GOVONI}
\address{Dipartimento di Astronomia, Universit\'a di
Bologna, Via Ranzani 1,\\
I-40127 Bologna, Italy \\
fgovoni@ira.cnr.it}

\author{LUIGINA FERETTI}
\address{Istituto di Radioastronomia CNR/INAF, Via Gobetti 101,\\
I-40129 Bologna, Italy \\
lferetti@ira.cnr.it}

\maketitle

\begin{history}
\received{(Day Month Year)}
\revised{(Day Month Year)}
\end{history}

\begin{abstract}
The existence of magnetic fields associated with the intracluster
medium in clusters of galaxies is now well established through
different methods of analysis.  Magnetic fields are investigated in
the radio band from studies of the rotation measure of polarized radio
galaxies and the synchrotron emission of cluster-wide diffuse sources.
Other techniques include X-ray studies of the inverse Compton emission
and of cold fronts and magneto hydrodynamic simulations.  We review
the main issues that have led to our knowledge on magnetic fields in
clusters of galaxies.  Observations show that cluster fields are at
the $\mu$G level, with values up to tens of $\mu$G at the center of
cooling core clusters.  Estimates obtained from different
observational approaches may differ by about an order of
magnitude. However, the discrepancy may be alleviated by considering
that the magnetic field is not constant through the cluster, and shows
a complex structure.  In particular, the magnetic field intensity
declines with the cluster radius with a rough dependence on the
thermal gas density. Moreover, cluster magnetic fields are likely to
fluctuate over a wide range of spatial scales with values 
from a few kpc up to
hundreds kpc.  Important information on the cluster field
are obtained by comparing the observational results with the
prediction from numerical simulations.  The origin of cluster magnetic
fields is still debated.  They might originate in the early Universe,
either before or after the recombination, or they could have been
deposited in the intracluster medium by normal galaxies, starburst
galaxies, or AGN. In either case, magnetic fields undergo significant
amplification during the cluster merger processes.
\end{abstract}


\section{Introduction}
Experience indicates that most of the matter in the Universe is
composed of ionized or partially ionized gas permeated by magnetic
fields. Celestial objects are magnetized and magnetic fields of
significant strength are found everywhere in the interstellar space,
and over small and very large scales, in the extragalactic universe.
In general, small compact objects have the largest magnetic field
strengths, while larger low-density objects have weaker magnetic fields.

The Earth has a bipolar field of about 0.5 G at its surface,
originating from an idealized current due to the charged fluid motion
going circularly in a ring inside the liquid molten metallic core.  On
the Jupiter surface the magnetic field is about 4 G, owing to the fast
Jupiter rotation.  In the interplanetary space of the solar system the
magnetic fields are of the order of 50 $\mu$G.

On the Sun, the magnetic field is of 10 G at the poles, while
localized sunspots on the surface near the equatorial zone of the Sun,
and more generally of a star, can have magnetic field strengths of
2000 G. In protostellar envelopes and protostars, fields are of $\sim$
1 mG.  A bipolar field is ``frozen'' into the gas of a star during the
contraction from a normal star to a degenerate star. It will remain
bipolar-shaped but its intensity will increase as r$^{-2}$, thus
magnetic fields of pulsars and neutron stars are of the order of
10$^{12}$ G, those of white dwarfs are around 10$^6$ G.

A widespread field of $\sim$ 5 $\mu$G, characterized by a spiral
shape, is present in the Galaxy. At the Galaxy nucleus, highly
organized filaments with strength of $\sim$ 1 mG are detected.
Fields in other spiral galaxies are of $\sim$ 10 $\mu$G on average,
with values up to $\sim$ 50 $\mu$G in starburst galaxies and $\sim$
30 $\mu$G in massive spiral arms.

Fields of $\sim$ $\mu$G are found in the radio emitting lobes of radio
galaxies. Fields of similar or weaker strength are detected in
the intracluster medium of clusters of galaxies, and in more rarefied
regions of the intergalactic space.  Upper limits of $\simeq 10^{-8}-
10^{-9}$ G have been obtained for the cosmological fields at large
redshift.

In this review large-scale magnetic fields in clusters of galaxies
will be analyzed.  In the last years the presence of cluster magnetic
fields has been unambiguously proven and the importance of their role
has been recently recognized. The study of cluster magnetic fields is
relevant to understand the physical conditions and energetics of the
intracluster medium. Cluster magnetic fields provide an additional
term of pressure and may play a role in the cluster dynamics.  They
couple cosmic ray particles to the intracluster gas, and they are able
to inhibit transport processes like heat conduction, spatial mixing of
gas, and propagation of cosmic rays.  They are essential for the
acceleration of cosmic rays and allow cosmic ray electron population
to be observed by the synchrotron radiation.

Despite many observational efforts to measure their properties, our
knowledge on cluster magnetic fields is still poor. Overviews on
observational and theoretical arguments can be found in the
literature\cite{kro94,bec96,grarub01,cartay02,wid02,gio04}.

The focus of this review is primarily observational, however, we
present the basic theory needed for the interpretation of the data.
We analyze some of the main issues that have led to our knowledge on
magnetic fields in clusters of galaxies and discuss some of their
limitations.  An outline of the review is as follows: In Sec. 2 we
summarize some general properties of clusters of galaxies. Sec. 3 is
devoted to theoretical background related to the detection of cluster
magnetic fields and to the estimate of their strengths.  We recall the
basic theory concerning synchrotron radiation, inverse Compton
radiation and Faraday rotation. These are the main observed features
which provide information on the cluster magnetic fields.  The
observational results of cluster magnetic fields through synchrotron
radio and inverse Compton hard X-ray emissions are described in
Secs. 4 and 5. In Sec. 6 we give the results obtained by analyzing
rotation measures of radio galaxies located within or behind clusters
of galaxies. In Sec. 7 we present cluster magnetic fields detected
through the study of cold fronts.  In Sec. 8 we report the evidence
for a radial decline of cluster magnetic fields. In Sec. 9 we discuss
how magnetic field values obtained with different approaches can be
reconciled.  In Sec. 10 we summarize the results of a numerical
technique which can significantly improve our interpretation of the
data and thus the knowledge of the strength and structure of magnetic
fields.  In Sec. 11 we briefly review the current knowledge on the
cluster magnetic field origin and amplification.

Throughout this paper we assume the $\lambda$CDM cosmology with
$H_0$ = 71 km s$^{-1}$Mpc$^{-1}$,
$\Omega_m$ = 0.3, and $\Omega_{\Lambda}$ = 0.7, unless stated
otherwise.

\section{Clusters of Galaxies}

Clusters of galaxies are the largest gravitationally bound systems in
the Universe. They appear at optical wavelengths as over-densities of
galaxies with respect to the field average density.  In addition to
the galaxies, they contain an intracluster medium (ICM) of hot
(T$\simeq 10^{8}$ K), low-density (n$_e \simeq 10^{-3}$ cm$^{-3}$)
gas, detected through its luminous X-ray emission (L$_X$$\simeq
10^{43}-10^{45}$ erg~s$^{-1}$), produced by thermal bremsstrahlung
radiation.

The visible galaxies and the ICM are important components of clusters,
however most of the cluster mass is in dark matter.  Although dark
matter has not been directly observed at any wavelength and its nature
remains unknown, X-ray and visible light observations provide clues to
its amount and distribution in clusters.

X-ray images, starting with the {\it Einstein} satellite and
continuing with {\it ROSAT} and {\it ASCA}, and now with {\it Chandra}
and {\it XMM-Newton}, provide a powerful technique to trace the global
cluster gravitational potential and to probe the dynamics, morphology
and history of clusters.  In the hierarchical scenario of the
structures formation, clusters of galaxies are formed by the
gravitational merger of smaller units e.g.  groups and
sub-clusters. Such mergers are spectacular events involving kinetic
energies as large as $\simeq 10^{64}$ ergs.  In these mergers a large
portion of energy is dissipated in the ICM, generating shock,
turbulence and bulk motions, and  heating it.
Substructure in the X-ray images as well as complex gas temperature
gradients are all signatures of cluster mergers.

A significant fraction of clusters of galaxies shows the X-ray surface
brightness strongly peaked at the center. This implies a high density,
and cooling times of the hot ICM within the inner $\simeq$ 100 kpc of
much less than the Hubble time. To maintain hydrostatic equilibrium,
an inward flow may be required.  X-ray observations with {\it
XMM-Newton} indicate no spectral evidence for large amounts of cooling
and condensing gas in the centers of galaxy clusters believed to
harbour strong cooling flows.  The cooling flow seems to be hindered
by some mechanism, whose nature is still debated.  Thus, there is no
consensus on the actual existence of material ``cooling'' and
``flowing''.  What is generally agreed upon is that cooling core
clusters are more dynamically relaxed than non cooling core clusters,
which often show evidence of cluster merger.

One  of the  most important  results obtained  with the  {\it Chandra}
satellite on clusters  of galaxies was the discovery  of sharp surface
brightness discontinuities  in the images of  merging clusters, called
``cold fronts''.   Initially, one might have  suspected these features
to be merger shocks but spectral measurements showed that these are a
new  kind of  structure.   These cold  fronts  are apparently  contact
discontinuities between the gas  which was in the cool core  of one of the
merging sub-clusters and the surrounding intracluster gas.
Cold fronts and merger shocks offer unique insights into the cluster
physics, including the determination of the gas bulk velocity, its
acceleration, the growth of plasma instabilities, the strength and
structure of magnetic fields and the thermal conductivity.

A precise physical description of the ICM necessitates also adequate
knowledge of the role of non-thermal components.  The most detailed
evidence for these phenomena comes from the radio observations. A
number of clusters of galaxies is known to contain wide diffuse
synchrotron sources (radio halos, relics and mini-halos) which have no
obvious connection with the cluster galaxies, but are rather
associated with the ICM.  The synchrotron emission of such sources
requires a population of $\approx$ GeV relativistic electrons and
cluster magnetic fields on $\mu$G levels.  An indirect evidence of the
existence of cluster magnetic fields is also derived from studies of
the Rotation Measure of radio galaxies located within or behind
clusters of galaxies.

A probe of the existence of a population of relativistic electrons in
the ICM is also obtained from the detection of non-thermal emission of
inverse Compton origin in the hard X-ray and possibly in the extreme
ultraviolet wavelengths.  The combination of the observed diffuse
radio and hard X-ray emissions from clusters of galaxies is used to
estimate the intracluster magnetic field strengths.

\section{Theoretical Background Related to Cluster Magnetic Fields}

\subsection{Synchrotron radiation}

The synchrotron emission is produced by the spiraling motion of
relativistic electrons in a magnetic field.  It is therefore the
easiest and more direct way to detect magnetic fields in astrophysical
sources.  The total synchrotron emission from a source provides an
estimate of the strength of magnetic fields while the degree of
polarization is an important indicator of the field uniformity and
structure.

\begin{figure}[th]
\centerline{\psfig{file=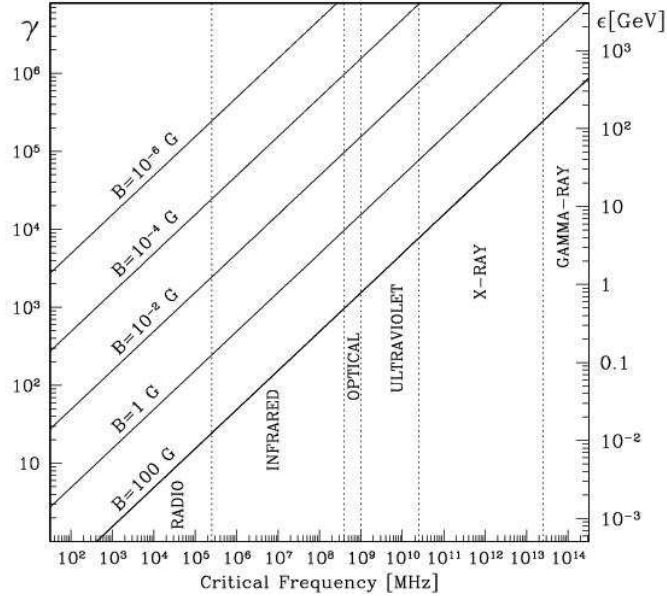,angle=0,width=9cm}}
\caption{Electron Lorentz factor $\gamma=\epsilon/m_ec^2$ 
(left-hand axis) and energy (right-hand axis) versus synchrotron
critical frequency for magnetic field strengths 
in the range $10^{-6}-100$ G ($\theta$=90\degrees). 
}
\label{fig1}
\end{figure}

An electron of energy $\epsilon=\gamma m_e c^2$ (where $\gamma$ is the
Lorentz factor) in a magnetic field $\vec{B}$ experiences a
$\vec{v}\times\vec{B}$ force that causes it to follow a helical path
along the field lines, emitting radiation into a cone of half-angle
$\simeq~\gamma^{-1}$ about its instantaneous velocity.  To the
observer, the radiation is essentially a continuum with a fairly
peaked spectrum concentrated near the critical frequency

\begin{equation}
\nu_c=c_1(Bsin\theta)\epsilon^2.
\label{sync}
\end{equation}
\noindent
The synchrotron power emitted by a relativistic electron is 

\begin{equation}
-\frac{d\epsilon}{dt}=c_2(Bsin\theta)^2\epsilon^2,
\end{equation} 
\noindent
where $\theta$ is the pitch angle between the electron velocity 
and the magnetic field direction while
$c_1$ and $c_2$ depend only on fundamental physical constants

\begin{equation}
c_1=\frac{3e}{4\pi m_e^3c^5},~~~~~~c_2=\frac{2e^4}{3m_e^4c^7}.
\end{equation}
\noindent 
In practical units:

\begin{equation}
{\nu_c}[MHz]\simeq 16.1\times10^6 (B_{[G]}sin\theta)(\epsilon_{[GeV]})^2
\end{equation} 
\noindent
$~~~~~~~~~~~~~~~~~~~~~~~~~~~~~~~~~~~~~\simeq 4.2(B_{[G]}sin\theta) \gamma^2$,

\begin{equation}
-\frac{d\epsilon}{dt}\left[\frac{erg}{s}\right]\simeq6.0\times10^{-9}(B_{[G]} sin\theta)^2 (\epsilon_{[GeV]})^2
\end{equation} 
$~~~~~~~~~~~~~~~~~~~~~~~~~~~~~~~~~~~~~\simeq1.6\times10^{-15}(B_{[G]} sin\theta)^2\gamma^2$.\\
\noindent
From Eq. \ref{sync}, it is derived that electrons of $\gamma \simeq
10^4$ in magnetic fields of $B\simeq 1$ G produce synchrotron
radiation in the optical domain, whereas electrons of $\gamma \simeq
10^5$ in magnetic fields of $B\simeq 10$ G radiate in the X-rays (see
Fig. \ref{fig1}).  Therefore at a given frequency, the energy (or
Lorentz factor) of the emitting electrons depends directly on the
magnetic field strengths.  The higher is the magnetic field strength,
the lower is the electron energy needed to produce emission at a given
frequency.  In a magnetic field of about $B\simeq 1$ $\mu$G, a
synchrotron radiation detected for example at $100$ MHz, is produced
by relativistic electrons with $\gamma \simeq 5000$.

For an homogeneous and isotropic population of electrons with a
power-law energy distribution, i.e. with the particle density between
$\epsilon$ and $\epsilon$+d$\epsilon$ given by
\begin{equation}
N(\epsilon)d\epsilon=N_0\epsilon^{-\delta}d\epsilon,
\label{powerlaw}
\end{equation}
\noindent
the total intensity spectrum,
in regions which are optically thin to their own radiation, varies as:
\begin{equation}
S(\nu)\propto \nu^{-\alpha},
\label{spectralind}
\end{equation}
\noindent
where the spectral index $\alpha=(\delta-1)/2$.
Below the frequency where the synchrotron emitting region becomes
optically thick, the total intensity spectrum
can be described by: 

\begin{equation} 
S(\nu)\propto\nu^{+5/2}.
\end{equation}

\noindent
The synchrotron emission radiating from a population of relativistic
electrons in a uniform magnetic field is linearly polarized.  In the
optically thin case, the degree of intrinsic linear polarization, for
a homogeneous and isotropic distribution of relativistic electrons
with a power-law spectrum as in Eq. \ref{powerlaw}, is:
\begin{equation}
P_{Int}=\frac{3\delta+3}{3\delta+7},
\label{pthin}
\end{equation}
\noindent
with the electric (polarization) vector perpendicular to
the projection of the magnetic field onto the plane of the sky.
For typical values of the particle spectral index, the intrinsic
polarization degree is $\sim$ $75-80$\%.
In the optically thick case:
\begin{equation}
P_{Int}=\frac{3}{6\delta +13}    
\label{pthick}
\end{equation}
\noindent
and the electric vector is parallel to the projected magnetic field.

In practice, the polarization degree detected in radio sources is much
lower than expected by the above equations.  A reduction in
polarization could be due to a complex magnetic field structure whose
orientation varies either with depth in the source or over the angular
size of the beam.  For instance, if one describes the magnetic field
inside an optically thin source as the superposition of two components,
one uniform $B_u$, the other isotropic and random $B_r$, the
observed degree of polarization can be approximated by\cite{bur66}:

\begin{equation}
P_{Obs}=P_{Int}{B_u^2 \over {B_u^2+B_r^2}}.
\label{poss}
\end{equation}

\noindent
A rigorous treatment of how the degree of polarization is
affected by the magnetic field configuration is presented by Sokoloff
et al.\cite{sok98,sok99}

\subsection{Equipartition magnetic fields 
derived from the synchrotron emission}

From the synchrotron emissivity it is not possible to derive
unambiguously the magnetic field value.  The usual way to estimate the
magnetic field strength in a radio source is to minimize its total
energy content $U_{tot}$\cite{pac70}.  The total
energy of a synchrotron source is due to the energy in relativistic
particles (U$_{el}$ in electrons and U$_{pr}$ in protons) plus the
energy in magnetic fields (U$_B$):
\begin{equation}
U_{tot}=U_{el}+U_{pr}+U_B.
\end{equation} 

\noindent
The magnetic field energy contained in the source volume $V$ is given by
\begin{equation} 
U_B = \frac{B^2}{8\pi}\Phi V,
\label{enb}
\end{equation}
\noindent
where $\Phi$ is the fraction of the source volume occupied by the magnetic
field (filling factor).
The electron total energy in the range $\epsilon_1-\epsilon_2$: 
\begin{equation}
 U_{el} =V\times \int_{\epsilon_{1}}^{\epsilon_{2}} N(\epsilon)\epsilon \, d\epsilon= VN_{0} \int_{\epsilon_{1}}^{\epsilon_{2}} \epsilon^{-\delta +1} \,d\epsilon 
\label{rif1}
\end{equation}
\noindent
can be expressed as a function of the synchrotron luminosity  $L_{syn}$:

\begin{equation}
 L_{syn} =V \times \int_{\epsilon_{1}}^{\epsilon_{2}} \left(-\frac{d\epsilon}{dt}\right)N(\epsilon) \, d\epsilon = c_2(Bsin\theta)^2 V N_{0} \int_{\epsilon_{1}}^{\epsilon_{2}} \epsilon^{-\delta +2} \,d\epsilon
\label{rif2}
\end{equation}

\begin{figure}[th]
\centerline{\psfig{file=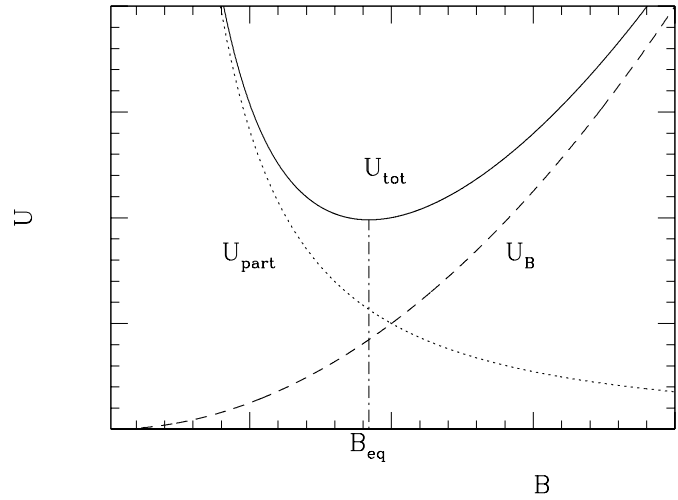,angle=0,width=9cm}}
\caption{ Energy content in a radio source (in arbitrary units): the
energy in magnetic fields is $U_B \propto$ B$^2$, the energy in
relativistic particles is $U_{part}=U_{el}+U_{pr} \propto$ B$^{-3/2}$.
The total energy content $U_{tot}$ is minimum when the contributions
of magnetic fields and relativistic particles are approximately equal
(equipartition condition). The corresponding magnetic field is
commonly referred to as equipartition value B$_{eq}$.}
\label{fig2}
\end{figure}

\noindent
by eliminating $V$$N_0$ 
and by writing $\epsilon_1$
and $\epsilon_2$ in terms of $\nu_1$ and $\nu_2$ (Eq. \ref{sync}):
\begin{equation}
 U_{el} = c_2^{-1}c_1^{1/2}\tilde{C}(\alpha,\nu_1,\nu_2){L_{syn}}{B^{-3/2}}=c_{12}(\alpha,\nu_1,\nu_2){L_{syn}}{B^{-3/2}},
\label{frequenze}
\end{equation}
\noindent
where $sin\theta$ has been taken equal to 1 and

\begin{equation}
\tilde{C}(\alpha,\nu_1,\nu_2)=\left(\frac{2\alpha - 2}{2\alpha - 1}\right) \frac{\nu_{1}^{(1-2\alpha)/2} - \nu_{2}^{(1-2\alpha)/2}}{\nu_{1}^{(1-\alpha)} - \nu_{2}^{(1-\alpha)}}.
\label{ctild}
\end{equation}
\noindent
The energy contained in the heavy particles, $U_{pr}$, can be related
to $U_{el}$ assuming: 
\begin{equation}
U_{pr} =kU_{el}.
\end{equation}
\noindent
Finally the total energy is obtained as a function of the magnetic field:
\begin{equation} 
U_{tot} = (1 + k)c_{12} L_{syn} B^{-3/2} + \frac{B^2}{8\pi}\Phi V.
\label{etot}
\end{equation}

\noindent
In order to obtain an estimate for the magnetic fields, it is
necessary to make some assumptions about how the energy is distributed
between the fields and particles.  A convenient estimate for the total
energy is represented by its minimum value (see Fig. \ref{fig2}).  
The condition of minimum
energy is obtained when the contributions of the magnetic field and
the relativistic particles are approximately equal:
\begin{equation} 
U_{B} =\frac{3}{4}(1+k)U_{el}.
\end{equation}
\noindent
For this reason the minimum energy is known as equipartition
value:

\begin{equation} 
U_{tot}(min) =\frac{7}{4}(1+k)U_{el}=\frac{7}{3}U_{B}.
\label{etotb} 
\end{equation}

\begin{table}[ht]
\tbl{Equipartition parametrization}  
{\begin{tabular}{@{}lcc@{}} \toprule 
 $\alpha$    & $\xi$($\alpha$, 10 MHz, 10 GHz)  & $\xi$($\alpha$, 10 MHz, 100 GHz) \\ \colrule
   0.0       &  1.43 $\times$ 10$^{-11}$  &  2.79 $\times$ 10$^{-11}$ \\
   0.1       &  9.40 $\times$ 10$^{-12}$  &  1.63 $\times$ 10$^{-11}$ \\
   0.2       &  6.29 $\times$ 10$^{-12}$  &  9.72 $\times$ 10$^{-12}$ \\
   0.3       &  4.29 $\times$ 10$^{-12}$  &  5.97 $\times$ 10$^{-12}$ \\
   0.4       &  2.99 $\times$ 10$^{-12}$  &  3.79 $\times$ 10$^{-12}$ \\
   0.5$^*$   &  2.13 $\times$ 10$^{-12}$  &  2.50 $\times$ 10$^{-12}$ \\ 
   0.6       &  1.55 $\times$ 10$^{-12}$  &  1.72 $\times$ 10$^{-12}$ \\
   0.7       &  1.15 $\times$ 10$^{-12}$  &  1.23 $\times$ 10$^{-12}$ \\
   0.8       &  8.75 $\times$ 10$^{-13}$  &  9.10 $\times$ 10$^{-13}$ \\
   0.9       &  6.77 $\times$ 10$^{-13}$  &  6.92 $\times$ 10$^{-13}$ \\
   1.0$^*$   &  5.32 $\times$ 10$^{-13}$  &  5.39 $\times$ 10$^{-13}$ \\
   1.1       &  4.24 $\times$ 10$^{-13}$  &  4.27 $\times$ 10$^{-13}$ \\
   1.2       &  3.42 $\times$ 10$^{-13}$  &  3.43 $\times$ 10$^{-13}$ \\
   1.3       &  2.79 $\times$ 10$^{-13}$  &  2.79 $\times$ 10$^{-13}$ \\
   1.4       &  2.29 $\times$ 10$^{-13}$  &  2.29 $\times$ 10$^{-13}$ \\
   1.5       &  1.89 $\times$ 10$^{-13}$  &  1.89 $\times$ 10$^{-13}$ \\
   1.6       &  1.57 $\times$ 10$^{-13}$  &  1.57 $\times$ 10$^{-13}$ \\
   1.7       &  1.31 $\times$ 10$^{-13}$  &  1.31 $\times$ 10$^{-13}$ \\
   1.8       &  1.10 $\times$ 10$^{-13}$  &  1.10 $\times$ 10$^{-13}$ \\
   1.9       &  9.21 $\times$ 10$^{-14}$  &  9.21 $\times$ 10$^{-14}$ \\
   2.0       &  7.76 $\times$ 10$^{-14}$  &  7.76 $\times$ 10$^{-14}$\\\botrule
\end{tabular}}
*for these values of $\alpha$ the constant defined in Eq. \ref{ctild} 
diverges, thus the corresponding parameters have been computed
by solving directly the integrals in Eqs. \ref{rif1} and  \ref{rif2}.
\end{table}

\noindent
The magnetic field for which the total energy content is minimum is:
\begin{equation}
B_{eq}=(6 \pi (1+k) c_{12} L_{syn}\Phi^{-1}V^{-1})^{2/7}. 
\label{beqfin}
\end{equation}
\noindent
The total minimum energy is:
\begin{equation}
U_{tot}(min) = c_{13}\left(\frac{3}{4\pi}\right)^{3/7} (1+k)^{4/7} \Phi^{3/7} V^{3/7} L_{syn}^{4/7},
\end{equation}
\noindent
and the total minimum energy density is:
\begin{equation}
u_{min}=\frac{U_{tot}(min)}{V \Phi} = c_{13}\left(\frac{3}{4\pi}\right)^{3/7}  (1+k)^{4/7}  \Phi^{-4/7} V^{-4/7} L_{syn}^{4/7}, 
\end{equation}
\noindent
where $c_{13}=0.921 c_{12}^{4/7}.$ 
The constants $c_{12}$ and $c_{13}$, 
depending on the spectral index and on the frequency range, 
are tabulated\cite{pac70} for cgs units.

By including the K-correction, assuming $\Phi$ = 1,
and expressing the parameters in commonly used units,
we can write the minimum energy density of a radio source
in terms of observed quantities:
\newpage
\begin{equation}
u_{min}\left[\frac{erg}{cm^3}\right]= \xi(\alpha,\nu_1,\nu_2) (1+k)^{4/7}  (\nu_{0[MHz]})^{4\alpha/7} (1+z)^{(12+4\alpha)/7}\times  
\label{intnu}
\end{equation}
$~~~~~~~~~~~~~~~~~~~~~~~~~~~~~~~\times(I_{0\left[\frac{mJy}{arcsec^2}\right]})^{4/7}(d_{[kpc]})^{-4/7}$,\\

\noindent
where $z$ is the source redshift, I$_0$ is the source brightness at
the frequency $\nu_0$, $d$ is the source depth, and the constant
$\xi$($\alpha, \nu_1,\nu_2$) is tabulated in Table 1 for the
frequency ranges: 10 MHz$-$10 GHz and 10 MHz$-$100 GHz.  I$_0$ can be
measured directly by the contour levels of a radio image (for
significantly extended sources), or can be obtained by dividing the
source total flux by the source solid angle.

The equipartition magnetic field is then obtained as:

\begin{equation}
B_{eq} = \left({{24\pi}\over{7}} u_{min}\right)^{1/2}.
\end{equation}

\noindent
One must be aware of the uncertainties inherent to this determination
of the magnetic field strength.  The value of $k$, ratio of the energy
in relativistic protons to that in electrons,  depends on the
mechanism of generation of relativistic electrons, which is so far
poorly known. Uncertainties are also related to the volume filling
factor $\Phi$. Values usually assumed in literature for clusters are
$k$ = 1 (or $k$ = 0) and $\Phi$ = 1.  Another parameter difficult to
infer is the extent of the source along the line of sight $d$.

In the standard approach presented above, the computation of
equipartition parameters is based on the integration of the
synchrotron radio luminosity between the two fixed frequencies $\nu_1$
and $\nu_2$ (Eq. \ref{frequenze} and followings). The electron energies
corresponding to these frequencies depend on the magnetic field value
(see Eq. \ref{sync}), thus the integration limits are variable in
terms of the energy of the radiating electrons.  The lower limit is
particularly relevant, owing to the power-law shape of the electron
energy distribution and to the fact that electrons of very low energy
are expected to be present.  If a low-energy cutoff in the particle
energy distribution is imposed, rather than a low-frequency cut-off in
the emitted synchrotron spectrum, the exponent 2/7 in Eq. \ref{beqfin}
should be replaced by 1/(3+$\alpha$), as pointed out by Beck \&
Krause\cite{beckra04}.  The equipartition quantities obtained
following this approach are presented by Brunetti et al.\cite{bru97}.
Indicating the electron energy by its Lorentz factor $\gamma$,
assuming that $\gamma_{min} << \gamma_{max}$, the new expression for
the equipartition magnetic field B$^{\prime}_{eq}$ in Gauss is 
(for $\alpha~>~0.5$):

\begin{equation}
B^{\prime}_{eq} \sim 1.1 ~ \gamma_{min}^{{1-2\alpha}\over{3+\alpha}}~B_{eq}^{{7}\over{2(3+\alpha)}},
\label{eqbru}
\end{equation}

\noindent
where B$_{eq}$ is the value of the equipartition magnetic field
obtained with the standard formulae by integrating the radio spectrum
between 10 MHz and 100 GHz.
If the equipartition
magnetic field obtained with the standard formulae is $\sim \mu$G,
the magnetic field derived considering electrons of $\gamma_{min}$ 
$\sim$ 100 is 2 to 5 times larger, using $\alpha$ in the range 
$0.75-1$.

\subsection{Inverse Compton radiation}

Relativistic electrons in a radiation field can scatter and transfer energy
to photons through the inverse-Compton (IC) effect.
This situation where the wave gains energy from the electron is the 
inverse of the usual Compton scattering.
The frequency of the scattered wave $\nu_{out}$ is related to that 
of the incident wave $\nu_{in}$ as:

\begin{equation}
\nu_{out} = { 4 \over 3 } \gamma^2 \nu_{in}.
\label{ic1}
\end{equation}
\noindent
In astrophysical applications, the IC plays a very important role
since the relativistic electron population responsible for synchrotron
emission scatters the ubiquitous 3K microwave background photons.  The
Planck function at T = 2.73 K peaks near a frequency of $\nu \sim 1.6\times
10 ^{11}$ Hz, therefore from Eq. \ref{ic1} the relativistic electrons of energy
$\gamma$ = 1000 $-$ 5000 will be responsible for IC emission in the X-ray domain,
respectively at $\sim$ $2\times 10^{17} - 5.3 \times 10^{18}$ Hz,
corresponding to $\sim$ 0.9 $-$ 22 keV. Microwave background 
photons are then turned into X-ray or gamma photons.  

Given that synchrotron and IC emission originate from the
same, assumed power-law, relativistic electron population
(Eq. \ref{powerlaw}), they share the same spectral index $\alpha$.
The spectral index relates to the index of the power-law electron 
energy density distribution as $\delta=2\alpha+1$, and to
the photon index of the IC emission as $\Gamma_X=\alpha + 1$.

\subsection{Magnetic fields derived from IC emission}

When the synchrotron radio and IC X-ray emission are produced by the 
same population of relativistic electrons (see Secs. 3.1 and 3.3),
the total synchrotron and IC powers are related.
The IC emissivity is proportional to the energy density in the photon field,
u$_{ph}$, which for the cosmological blackbody radiation is $\sim 5\times
10^{-13}$ (1+z)$^4$ erg cm$^{-3}$, whereas the synchrotron emissivity is
proportional to the energy density in the magnetic field, u$_{B}$ =
B$^2$/8$\pi$. This leads to a simple proportionality between
synchrotron and IC luminosities:

\begin{equation}
{L_{syn} \over L_{IC}} \propto {u_B \over u_{ph}}. 
\end{equation}
\noindent
Combining the standard formulae of the synchrotron and Compton
emission mechanisms, the radio and HXR detections directly yield some
of the basic properties of the magnetic field. Following Blumenthal
and Gould\cite{blugou70}, the synchrotron flux at the
radio frequency $\nu_r$ and the IC X-ray flux at frequency
$\nu_x$ are (all parameters in cgs units):

\begin{equation}
S_{syn(\nu_r)}= 1.7\times 10^{-21}{{VN_0}\over{4\pi D^2}} a(\delta) B^{1+\alpha} \left({{4.3\times 10^6}\over{\nu_r}}\right)^{\alpha}, 
\end{equation}

\begin{equation}
S_{IC(\nu_x)}= 4.2\times 10^{-40}{{VN_0}\over{4\pi D^2}} b(\delta) T^{3 + \alpha}(1+z)^{3+\alpha} \left({{2.1\times 10^{10}}\over{\nu_x}}\right)^{\alpha},
\end{equation}
\noindent
where the functions a($\delta$) and b($\delta$) are tabulated in Table 2, 
V is the emission volume, and D the source distance. 
From the ratio between the X-ray and radio fluxes,
one derives an estimate of the total magnetic field, 
averaged over the emitting volume.

To obtain a formula for practical use, we first relate the
monochromatic X-ray flux S$_{IC(\nu_x)}$ to the flux S$_{IC(E_1-E_2)}$
integrated over the energy interval $E_1 - E_2$,
as this is the parameter usually measured from observations: 

\begin{equation}
S_{IC(E_1-E_2)} \propto S_{IC(\nu_x)}{{E_2^{1-\alpha}-E_1^{1-\alpha}}\over{1-\alpha}}(\nu_x)^{\alpha}.
\end{equation}

\noindent
We also substitute the radiation temperature T = 2.7 K at z = 0, and we 
compute the constants for commonly used units.
We obtain the magnetic field as: 

\begin{equation}
 (B[\mu G])^{1+\alpha} = h(\alpha){S_{syn(\nu_r)}[Jy]\over
{S_{IC(E_1-E_2)}[erg s^{-1} cm^{-2}]}}(1+z)^{3+\alpha}(0.0545 \nu_r[MHz])^{\alpha}\times
\label{campoic}
\end{equation}
$~~~~~~~~~~~~~~~~~~~~~\times(E_2[keV]^{1-\alpha}-E_1[keV]^{1-\alpha})$,

\noindent
where the function h($\alpha$) is tabulated in Table 2. 
For $\alpha=1$, the above formula becomes
\begin{equation}
(B[\mu G])^{2} = 10^{-16}{S_{syn(\nu_r)[Jy]}\over{S_{IC(E_1-E_2)}[erg s^{-1} cm^{-2}]}}(1+z)^{4} \frac{\nu_r}{[MHz]} \left(ln {{E_2}[keV] \over {E_1}[keV]}\right).
\label{bica2}
\end{equation}

\begin{table}[ht]
\tbl{Inverse Compton parametrization} 
{\begin{tabular}{ccccc} \toprule 
$\alpha$ & $\delta$ & a($\delta$) & b($\delta$) & h($\alpha$)  \\ \colrule
 0.0 & 1   & 0.283 & 3.20 & 1.32 $\times$ 10$^{-16}$ \\ 
 0.1 & 1.2 & 0.209 & 3.42 & 2.13 $\times$ 10$^{-16}$ \\
 0.2 & 1.4 & 0.164 & 3.73 & 3.31 $\times$ 10$^{-16}$ \\
 0.3 & 1.6 & 0.136 & 4.12 & 5.06 $\times$ 10$^{-16}$ \\
 0.4 & 1.8 & 0.117 & 4.62 & 7.71 $\times$ 10$^{-16}$ \\
 0.5 & 2.0 & 0.103 & 5.25 & 1.19 $\times$ 10$^{-15}$ \\
 0.6 & 2.2 & 0.093 & 6.03 & 1.89 $\times$ 10$^{-15}$ \\
 0.7 & 2.4 & 0.086 & 7.00 & 3.17 $\times$ 10$^{-15}$ \\    
 0.8 & 2.6 & 0.081 & 8.20 & 5.95 $\times$ 10$^{-15}$ \\
 0.9 & 2.8 & 0.077 & 9.69 & 1.48 $\times$ 10$^{-14}$ \\   
 1.0 & 3.0 & 0.074 & 11.54 &  see Eq. \ref{bica2} \\
 1.1 & 3.2 & 0.072 & 13.85 & $-$2.24 $\times$ 10$^{-14}$ \\   
 1.2 & 3.4 & 0.071 & 16.74 & $-$1.37 $\times$ 10$^{-14}$ \\   
 1.3 & 3.6 & 0.071 & 20.35 & $-$1.12 $\times$ 10$^{-14}$ \\
 1.4 & 3.8 & 0.072 & 24.89 & $-$1.02 $\times$ 10$^{-14}$ \\
 1.5 & 4.0 & 0.073 & 30.62 & $-$9.88 $\times$ 10$^{-15}$  \\
 1.6 & 4.2 & 0.075 & 37.87 & $-$9.96 $\times$ 10$^{-15}$ \\
 1.7 & 4.4 & 0.076 & 47.07 & $-$1.03 $\times$ 10$^{-14}$\\
 1.8 & 4.6 & 0.079 & 58.78 & $-$1.09 $\times$ 10$^{-14}$ \\
 1.9 & 4.8 & 0.083 & 73.74 & $-$1.16 $\times$ 10$^{-14}$ \\
 2.0 & 5.0 & 0.087 & 92.90 & $-$1.25 $\times$ 10$^{-14}$  \\ \botrule
\end{tabular}}
\end{table}

\noindent
The difficulties related to this method
are essentially due to the limitations of present X-ray observations
in the hard X-ray domain and to the problem of  distinguishing
between the non-thermal and the thermal X-ray emission. 
When the IC X-ray emission is not detected from a radio emitting
region, only lower limits to the magnetic fields can be derived.

\subsection{Faraday Rotation effect}

The Faraday rotation effect appears during the propagation
of electromagnetic waves in a magnetized plasma.
A linearly polarized wave can be decomposed into
opposite-handed circularly polarized components.
The right-handed and left-handed circularly 
polarized waves propagate with different phase velocities
within the magneto-ionic material.
This effectively rotates the plane
of polarization of the electromagnetic wave.

According to the dispersion relation, for a wave of 
angular frequency $\omega$ ($\omega=2\pi \nu$),
the refractive index of a magnetized dielectric medium 
can take two possible values:

\begin{equation}
n_{L,R}= \left(1-{{\omega_p^2}\over{\omega^2\pm \omega\Omega_e}}\right)^{1/2},
\label{indexrefr}
\end{equation}
\noindent
where $\omega_p$ = $({{4\pi n_e e^2}\over{m_e}})^{1/2}$ is the
plasma frequency, and $\Omega_e$ = ${{eB}\over{m_e c}}$ is the
cyclotron frequency.

In the context of the study of cluster magnetic fields
we are interested in the Faraday rotation of radio sources 
in the background of the cluster or in the cluster itself.
The radio frequencies dominate the values of $\omega_p$ and 
$\Omega_e$ obtained for typical magnetic 
fields ($B\simeq1\mu$G) and gas densities
($n_e\simeq10^{-3}$ cm$^{-3}$) in the ICM.
In the limit  $\omega >> \Omega_e$, Eq. \ref{indexrefr}
can be approximated as:

\begin{equation}
n_{L,R} \approx 1-{1 \over 2}{{\omega_p^2}\over{(\omega^2\pm\omega\Omega_e)}},
\label{indexrl}
\end{equation}
\noindent
thus the difference in time of the two
opposite handed waves to travel a path length $dl$ results:

\begin{equation}
\Delta t  \approx {{\omega_p^2 \Omega_e dl}\over {c \omega^3}} =
{{4 \pi e^3}\over{\omega^3 m_e^2 c^2}}n_e B dl 
\label{deltat}
\end{equation}
\noindent
and the phase difference between the two signals is
$\Delta \phi= \omega \Delta t$.
Therefore, traveling along the cluster
path length L, the 
intrinsic polarization angle $\Psi _{Int}$
will be rotated by an angle $\Delta \Psi$=$\frac{1}{2} \Delta \phi$, 
resulting: 

\begin{equation}
\Psi_{Obs}(\lambda) =\Psi _{Int}+\Delta \Psi=\Psi _{Int}+\frac{e^3\lambda^2}{2\pi m_e^2c^4}\int\limits_0^L n_e(l) B_{\|}(l)dl,
\end{equation}
\noindent
where $B_{\|}$ is the component of the magnetic field
along the line of sight.
$\Psi_{Obs}$ is usually written in terms of the rotation measure, RM:
\begin{equation}
\Psi_{Obs}(\lambda)=\Psi _{Int}+ \lambda^2RM,
\label{fit}
\end{equation}
\noindent
where: 
\begin{equation}
RM=\frac{e^3}{2\pi m_e^2c^4}\int\limits_0^L n_e(l) B_{\|}(l)dl.
\label{rmphys}
\end{equation}
\noindent
In practical units:
\begin{equation}
 RM\left[\frac{rad}{m^2}\right]= 812 \int\limits_0^Ln_e[cm^{-3}]B_{\|}[\mu G]dl[kpc].
\end{equation}
\noindent
By convention, RM is positive (negative) for a magnetic field
directed toward (away from) the observer.

The position angle of the polarization plane $\Psi_{Obs}$ is an observable
quantity, therefore, the RM of radio sources can be derived by
a linear fit to Eq. \ref{fit}.
In general, the position angle must be measured at three or more 
wavelengths in order to determine RM accurately and remove the
$\Psi_{Obs}=\Psi_{Obs}\pm n \pi$ ambiguity.

\subsubsection{Depolarization due to Faraday rotation}

The term depolarization indicates a decrease of the polarization
percentage, either at a given frequency, or when comparing two
different frequencies.  In a radio source the observed degree of
polarization intensity, $P_{Obs}(\lambda)$, can be significantly
lower with respect to the intrinsic value, $P_{Int}$, if
differential Faraday rotation occurs.  The Faraday rotation can induce
a depolarization of the observed radiation in different circumstances.

External depolarization is induced by the limitations of the instrumental
capabilities. Beamwidth depolarization is due to the presence of
fluctuations in the foreground screen within the observing beam:
unresolved density or magnetic field inhomogeneities of the media
through which the radiation propagates induce unresolved spatial
variation in the Faraday rotation measure and hence beam
depolarization.  In addition, bandwidth depolarization occurs when a
significant rotation of the polarization angle of the radiation is
produced across the observing bandwidth.

Internal depolarization is due to the spatial extent of the source itself
and occurs even if the intervening media are completely homogeneous.
Along the line of sight, the emission from individual electrons within
a source arises from different depths and suffers different Faraday
rotation angles due to the different path length.  For the total
radiation emitted by the source, this results in a reduction of the
observed degree of polarization.
In the case that the Faraday effect originates entirely within the
source, when the source can be represented by an homogeneous optically thin
slab, the degree of polarization varies as\cite{bur66}:
\begin{equation} 
P_{Obs}(\lambda) =P_{Int}{{sin 
(RM^{\prime}\lambda^2)}\over{RM^{\prime}\lambda^2}},
\label{depo} 
\end{equation} 

\noindent
where $RM^{\prime}$ is the internal Rotation Measure through the
depolarizing source.  If a value of the Rotation Measure $RM_{Obs}$ is
derived observationally from the rotation of the polarization angle,
then $RM^{\prime}$ in the above equation is =
2~$RM_{Obs}$\cite{sok98,sok99}. Indeed the observed rotation is the
average of the full rotation occurring across the source, thus it is
1/2 of the total back-to-front rotation.
 
To distinguish between the external and  internal depolarization,
very high resolution and sensitive polarization data at 
multiple frequencies are needed.
The key difference between them is 
that internal depolarization should be correlated with 
the Faraday rotation measure, therefore regions with small
RM should exhibit very little depolarization. 
Instead, the external beam depolarization, due to gradients in the RM, 
should not be correlated with the amount of the RM but with the
amount of the RM gradient.

\subsubsection{Interpretation of the cluster RM data}

RM data of radio sources in the background of clusters or in the
clusters themselves, together with a model for the intracluster gas
density distribution, can provide important information on the cluster
magnetic field responsible for the Faraday effect.
The Faraday effect of an external screen containing a gas with a
constant density and a uniform magnetic field produces no
depolarization and a rotation of the polarization angle proportional
to $\lambda^2$\rmm, with:
\begin{equation} 
\langle RM \rangle = 812 B_{\parallel} n_e L.
\end{equation}
\noindent
The existence of small-scale magnetic field structures
produce both rotation of the polarization angle and
depolarization.

The effect of Faraday rotation from a tangled magnetic field has been
analyzed by several
authors\cite{sok98,sok99,lawden82,tri91,fer95,fel96}, in the simplest
approximation that the magnetic field is tangled on a single scale
$\Lambda_{c}$.  In this ideal case, the screen is made of
cells of uniform size, electron density and magnetic field strength,
but with a field orientation at random angles in each cell. The
observed RM along any given line of sight is then generated by a
random walk process involving a large number of cells of size
$\Lambda_{c}$.  The distribution of the RM is Gaussian with \rmm =
0, and variance given by:
\begin{equation}
 \sigma_{RM}^{2}= \langle {RM^{2}} \rangle = 812^{2} \Lambda_{c} \int ( n_{e} B_{\|})^{2}dl~.
\label{sigmarndwalk}
\end{equation}
\noindent
In this formulation, by considering a 
density distribution which follows a $\beta$-profile\cite{cavfus76}:

\begin{equation} 
n_e(r)=n_0(1+r^2/r_c^2)^{-3\beta/2},
\label{king}
\end{equation}

\noindent
the following relation for the RM dispersion as a function 
of the projected distance from the cluster center, 
$r_{\perp}$, is obtained by integrating Eq. \ref{sigmarndwalk}:

\begin{equation} 
\sigma_{RM}(r_{\perp})= {{K B n_{0}  r_c^{1/2} \Lambda_{c}^{1/2} }
 \over {(1+r_{\perp}^2/r_c^2)^{(6\beta -1)/4}}} \sqrt {{\Gamma(3\beta-0.5)}\over{\Gamma(3\beta)}},
\label{felten}
\end{equation}

\noindent
where $\Gamma$ is the Gamma function. 
The constant $K$ depends on the integration path over 
the gas density distribution:
$K$ = 624, if the source lies completely
beyond the cluster, and $K$ = 441 if the source is halfway through
the cluster.

Therefore, since the density profile of the ICM can be obtained
by X-ray observations, the cluster magnetic field strength
can be estimated by measuring $\sigma_{RM}$ from
spatially resolved RM images of radio sources if $\Lambda_{c}$
is inferred or is known.

\section{Diffuse Radio Emission in Clusters of Galaxies}

The presence of magnetic fields in clusters is directly demonstrated
by the existence of large-scale diffuse synchrotron sources, that have
no apparent connection to any individual cluster galaxy and are
therefore associated with the ICM.  These radio sources have been
classified as radio halos, relics and mini-halos depending on their
morphology and location.  Radio halos, relics and mini-halos are not a
common phenomenon in clusters and until recently they were known to
exist in only a handful of clusters of galaxies\cite{fergio96}. 
This was a consequence of
the fact that they show low surface brightness, large size and steep
spectrum, thus they are difficult to reveal.  Extended, low-brightness
structures are most easily detected with filled-aperture telescopes,
but the low resolving power of single-dish radio telescopes 
increases problems with
confusion and can create an apparent wide source from a blend of weak
discrete radio sources.  Observations made with interferometers have
the angular resolution necessary to separate the individual radio
galaxies, but generally lack information from short spacing, thus
hindering the detection of extended low-brightness structures.

The number of clusters with known diffuse sources has increased in the
last few years to around 50,
thanks to the improved sensitivity of radio telescopes and the
existence of deep surveys.  New halo and relic candidates were found
from searches in the NRAO VLA Sky Survey (NVSS\cite{con98}) 
by Giovannini et al.\cite{gio99}, in the Westerbork Northern 
Sky Survey (WENSS\cite{ren97}) by Kempner and Sarazin\cite{kemsar01}, 
in the Sidney University Molonglo Sky Survey (SUMSS\cite{boc99})
by Hunstead et al.\cite{hun99} and in the survey of the Sharpley 
Concentration by Venturi et al.\cite{ven00}.

The presence of these large regions of diffuse synchrotron emission
reveals a large-scale distribution of $\approx$ GeV relativistic electrons 
radiating in $\sim$ $\mu$G magnetic fields, in the ICM.

\subsection{Radio halos}

\begin{figure}[th]
\centerline{\psfig{file=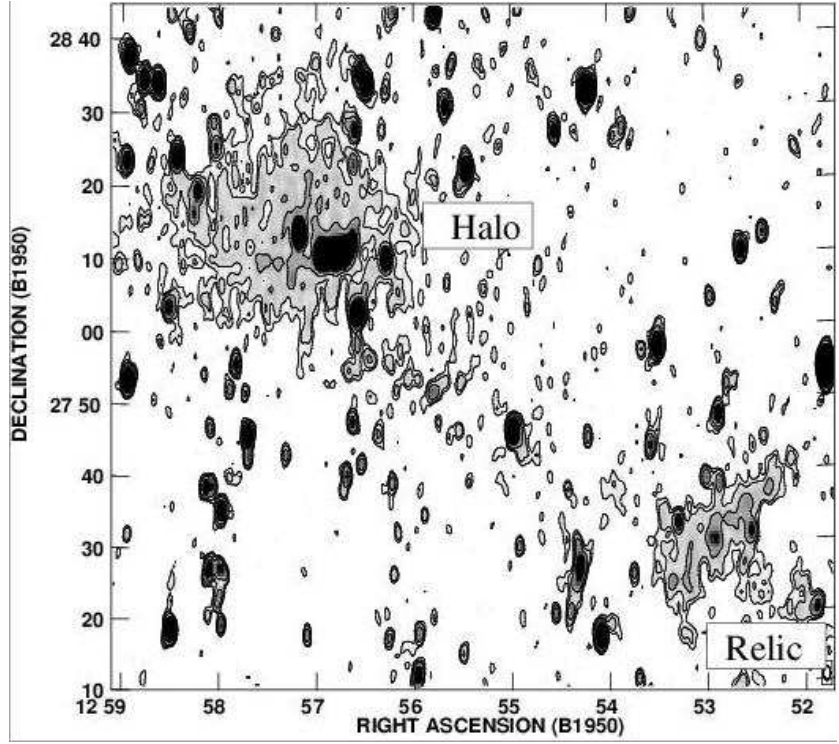,angle=0,width=11cm}}
\caption{WSRT radio image$^{34}$ of the Coma
cluster at 90 cm with a resolution (FWHM) of 55\arcsec$\times$
125\arcsec (RA $\times$ DEC).  The cluster center is approximately
located at the position RA$_{1950}$ = 12$^h$ 57$^m$ 24$^s$,
DEC$_{1950}$ = 28\degrees 15\arcmin 00\arcsec.  The radio halo Coma C
is at the cluster center, the radio relic 1253+275 is at the cluster
periphery.  The gray-scale range display total intensity emission from
2 to 30 mJy/beam whereas contour levels are at 3, 5, 10, 30, 50
mJy/beam.  The Coma cluster is at a redshift of 0.023, such that
1\arcsec~ is 0.46 kpc.
}
\label{fig3}
\end{figure}

Cluster radio halos are the most spectacular expression of clusters
non-thermal emission. They permeate the cluster centers with size of
more than a Mpc, showing low surface brightnesses ($\simeq$
10$^{-6}$Jy arcsec$^{-2}$ at 21 cm) and steep spectra ($\alpha$ \gtsim
1).
A typical example is Coma C, the halo source in the Coma cluster,
which was first shown to be diffuse by Willson\cite{wil70} and mapped
later at various radio wavelengths by several 
authors\cite{jaf76,val78,han79,sch87,hen89,kim90,gio93,dei97,ens99a,thi03}.
In Fig. \ref{fig3} we show a
radio image at 90 cm of the Coma cluster\cite{gio93}, 
obtained with the Westerbork
Synthesis Radio Telescope (WSRT).  
The integrated spectrum of Coma C is $\alpha \simeq$ 1.3, with a
steepening at high frequencies\cite{thi03}. The spectral
index distribution shows a radial
decrease from $\alpha \sim$ 0.8 at the cluster center, to $\alpha
\sim$ 1.8 at about 15\arcmin~ from the center\cite{gio93}.

Studies of several radio halos and of their hosting clusters
have been recently performed, thus improving the knowledge of
the characteristics and physical properties of this class of radio
sources.  Radio halos have been studied in X-ray luminous clusters
such as A2163\cite{fer01} and 1E0657-57\cite{lia00}, and 
in distant clusters, such as A2744\cite{gov01a} (z = 0.308), 
and CL0016+16\cite{giofer00} (z = 0.5545). The
latter is the most distant cluster with a radio halo known so far.
Halos of small size, i.e. $\sim$ 500-600 kpc
have also been detected in some cases
(e.g. in A2218\cite{giofer00} and in  A3562\cite{ven03}).

No polarized flux has been detected so far in radio halos.  In the
Coma cluster, the upper limit to the fractional polarization of Coma C
is of $\sim$ 10\% at 1.4 GHz\cite{gio93}.  Upper limits of $\sim$
6.5\% and of $\sim$ 4\% have been obtained for the two powerful radio
halos in A2219\cite{bac03} and A2163\cite{fer01}, respectively.
Also, no significant polarization is reported for
1E0657-57\cite{lia00}. The interpretation of these low polarization
levels is that the thermal gas has become mixed with the relativistic
plasma, thus internal depolarization occurs within the radio emitting
plasma. In addition, the magnetic field may be disordered on scales
smaller than the observing beam, thus producing significant beam
depolarization.

Due to their low surface brightness, radio halos have been studied so
far with low spatial resolution. This prevents a detailed
investigation of the small-scale magnetic field geometry and
intensity.  Using minimum energy assumptions (see Sec. 3.2),
it is possible to estimate an equipartition magnetic field strength
averaged over the entire halo volume, i.e. on scales as large as $\sim
1$ Mpc.  The derived minimum energy densities in halos are of the
order of $10^{-14}-10^{-13}$ erg cm$^{-3}$, i.e. much lower than the
energy density in the thermal gas.  These calculations typically
assume equal energy in relativistic protons and electrons (k = 1), a
volume filling factor $\Phi$ = 1, a low frequency cut-off of 10 MHz, and
a high frequency cut-off of 10 GHz.  The corresponding equipartition
magnetic field strengths range from $\simeq$ 0.1 to 1 $\mu$G.
In the Coma cluster, a minimum energy density of 1.9$\times$10$^{-14}$
erg cm$^{-3}$ is derived from the radio data in Coma C.  The
corresponding\cite{gio93} equipartition magnetic field is 0.45 $\mu$G.
In the equipartition approximation, a homogeneous cluster magnetic
field is assumed, but this is probably a too simple picture.
Important clues about a radial decrease of the magnetic field strength
in clusters of galaxies are given in Sec. 8.

Several models for the origin of the relativistic radiating electrons
in the ICM have been proposed (see, e.g., recent 
reviews\cite{ens02,sar02,bru03,pet03} and references
therein). These can be basically divided in two different scenarios:\\

$\bullet$ {\it Primary electron} models\cite{jaf77,rol81} in which
relativistic electrons are injected in the ICM from AGN activity
(quasar, radio galaxies, etc.) and/or from star formation in galaxies
(supernovae, galactic winds, etc.).  The radiative lifetime of the
relativistic electrons is relatively short ($\sim 10^{7-8}$
yrs). Therefore models involving a primary origin of the relativistic
electrons require continuous injection processes and/or reacceleration
processes in order to explain the presence of diffuse non-thermal
emission out to Mpc scales.  Electrons are likely reaccelerated in the
gas turbulence\cite{sch87,bru01,ohn02,fuj03} or in
shocks\cite{sar99,fujsar01,kes04}, although the efficiency of the latter
process is debated\cite{gabbla03,ryu03}.\\

$\bullet$ {\it Secondary electron}
models\cite{den80,blacol99,dolens00} in which cosmic ray
electrons result as secondary products of hadronic collisions between
relativistic protons and ICM thermal protons. The relativistic protons
in the ICM have lifetimes of the order of the Hubble time.  Thus
they are able to travel a large distance from their source before
they release their energy. In this way, electrons are produced through
the whole cluster volume and do not need to be reaccelerated.  The
production of relativistic electrons by secondary models predict large
gamma-ray fluxes from neutral pion decay which could be tested by
future gamma-ray missions.\\

On the observational side, it is possible to draw
some of the general characteristics of radio halos and derive
correlations with other cluster properties:\\ 

\noindent
i) Halos are typically found in clusters with significant
substructure and deviation from spherical symmetry
in the  X-ray morphology\cite{fer99a,buo01}.
This is confirmed by the high resolution X-ray data 
obtained with {\it Chandra} 
and {\it XMM}\cite{marvik01,mar02,mar03a,mar03b,kemdav04,gov04,hen04}.  
In addition to the
distorted X-ray morphology, all the clusters with halos exhibit 
strong gas temperature gradients. Some clusters show a spatial
correlation between the radio halo brightness and the hot gas regions,
although this is not a general feature\cite{gov04}.\\
ii) In a number of well-resolved clusters, a point-to-point
spatial correlation is observed between the radio brightness of the
halo and the X-ray brightness as detected by {\it ROSAT}\cite{gov01b}.  
This correlation is visible e.g. in A2744 also in the {\it Chandra} high
resolution data\cite{kemdav04}.  \\
iii) Halos are present in rich clusters, characterized by high
X-ray luminosities and temperatures\cite{gio02}.  
The percentage of clusters with
halos in a complete X-ray flux-limited sample (that includes systems
with $L_X> 5\times 10^{44}h_{50}^{-2}$ erg s$^{-1}$ 
in the $0.1-2.4$ keV band) is
$\simeq 5$\%.  The halo fraction increases with the X-ray luminosity,
to $\simeq 33$\% for clusters with $L_X> 10^{45}h_{50}^{-2}$ erg s$^{-1}$.\\
iv) The radio power of a halo  strongly
correlates with the cluster luminosity\cite{lia00,bac03,fer02} the
gas temperature\cite{lia00,col99}, and the
total mass\cite{gov01a}.\\

Therefore the available data suggest that radio halos seem to be
strictly related to the X-ray properties of the host clusters and to
the presence of cluster merger processes, which can provide the energy
for the electron reacceleration and magnetic field amplification on
large scales.
From energetic grounds, mergers can indeed supply enough
kinetic energy for the maintenance of a radio halo, as first
suggested by Harris et al.\cite{har80}.

The observed link between radio halos and cluster mergers is 
in favor of primary electron models.  These are also supported by
the high frequency steepening of the integrated radio spectra 
(e.g. in Coma C\cite{thi03}) and by the radial 
steepening of the two-frequency spectra
in Coma C\cite{gio93},  A665\cite{fer04} and A2163\cite{fer04}.
These spectral behaviors can be easily reproduced by models invoking
reacceleration of particles. On the contrary, they are difficult to
explain by models considering secondary electron populations.

\subsection{Relics}

Radio relics are a class of diffuse sources typically located near the
periphery of the cluster.  Unlike the halos, they show an elongated or
irregular shape and are strongly polarized.
The prototype of this class is 1253+275, in the Coma cluster (see
Fig. \ref{fig3}), first classified by Ballarati et al.\cite{bal81}.
The polarization of 1253+275 at 20 cm is $25-30$\%\cite{thi03,gio91}.
The magnetic field is oriented along the source major axis and the
magnetic field strength derived from minimum energy arguments of the
synchrotron emitting plasma is 0.55 $\mu G$.

A complex radio emission is detected in A2256\cite{brifom76,rot94},
which contains several head-tail radio galaxies, two large regions of
diffuse radio relic emission and a central radio halo. The relics are
highly polarized\cite{claens01}, with the linear fractional polarization
at 20 cm above 30\% for the majority of the region,
reaching values up to 50\%.  The intrinsic magnetic field direction
reveals that there is a large-scale order to the fields, and it
appears to trace the bright filaments in the relics.

In addition to Coma and A2256, other clusters
present both a central halo and a peripheral relic,
e.g. A2255\cite{bur95,fer97}, A1300\cite{rei99}, A2744\cite{gov01a} and
A754\cite{bac03,kas01}.
A spectacular example of two almost symmetric relics in the same
cluster is found in A3667\cite{rot97,joh03}.

A puzzling relic source is 0917+75\cite{giofer00,har93}, located at 5
to 8 Mpc from the centers of the closest clusters (A762, A786, A787),
thus not unambiguously associated with any of them. It has a high
fractional polarization (up to 48\% at 20 cm and 60\% at 6 cm for the
brightest parts of the source) and the magnetic field direction
appears to be coherent over scales of at least several hundreds kpc.

\begin{figure}[th]
\centerline{\psfig{file=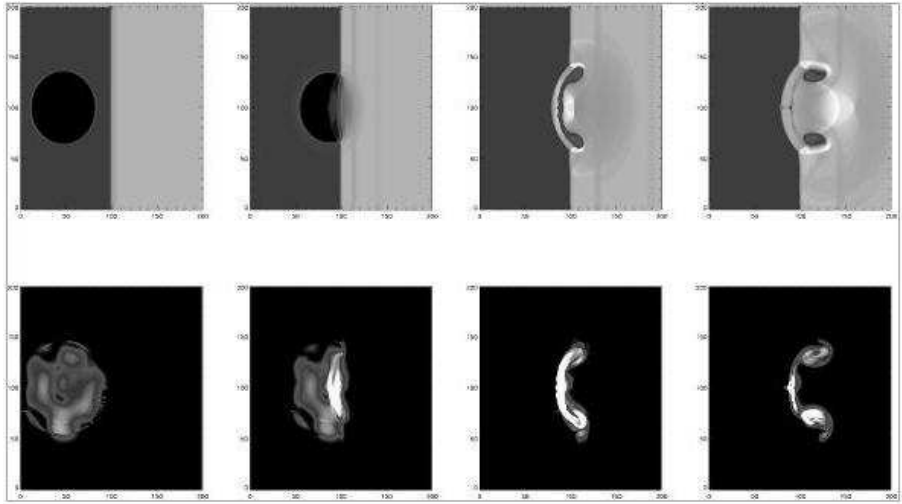,angle=0,width=12cm}}
\caption{ 
3-D MHD simulations$^{91}$
of electrons reaccelerated by compression of existing 
cocoons of radio plasma that traverses a shock wave.
In figure is shown the passage of a radio cocoon through a shock wave and the
consequent evolution of the gas density (top) and magnetic field energy 
density (bottom).
}
\label{fig4}
\end{figure}

The equipartition magnetic fields in the relics,  
computed with standard assumptions
(k = 1, $\Phi$ = 1, $\nu_1$ = 10 MHz, $\nu_2$ = 10 GHz),
are in the range  $0.5-2$ $\mu$G\cite{gov01a,gio91,ens98}. 
We note, however, that they refer to
regions where the cluster magnetic fields are expected to be
compressed (see below), thus they are not indicative of the overall
magnetic field intensity in the peripheral cluster regions.

In recent years, there has been increasing evidence that the relics
are related to ongoing merger events.
It has been suggested that relics 
result from first order Fermi acceleration of relativistic particles in
shocks produced during cluster merger events\cite{kes04,ens98,roe99a,min01}.
En{\ss}lin and Br\"uggen\cite{ensbru02} presented 3-D magneto
hydrodynamic (MHD) simulations
of electrons reaccelerated by adiabatic compression\cite{ensgop01}
of existing
cocoons of radio plasma that traverses a shock wave.  The passage of 
a radio cocoon through a shock wave and the consequent evolution of the
gas density and magnetic field energy density can be seen in
Fig. \ref{fig4}.  This model is consistent with the relic elongated
structure almost perpendicular to the merger axis.  Moreover, the
derived maps obtained in the simulations reproduce very well the
filamentary structure seen in relic sources at high
resolution\cite{sle01}.

The observed high polarization fraction of the relics should result
from the compression of the wave, which aligns unordered magnetic
fields with the shock front.  Due to the low gas density, a
low Faraday effect is expected at the cluster periphery, even in the
presence of a tangled magnetic field.

\subsection{Mini-halos}

There are a few clusters where the relativistic electrons can be
traced out quite far from the central galaxy, forming what is called
mini-halo. Mini-halos are diffuse steep-spectrum radio sources,
extended on a moderate scale (up to $\simeq$ 500 kpc), surrounding a
dominant radio galaxy at the cluster center.  Unlike radio halos and
relics, mini-halos are not tied to on-going merger events in clusters,
as they are typically found at the center of cooling core, i.e
relaxed, clusters.  The prototype example of a mini-halo is at the
center of the Perseus cluster.  The size is $\sim$ 450 kpc, with no
significant polarization\cite{sij93,bur92}.  The strong polarized
emission, detected through the entire cluster at 92 cm, at a 
Faraday depth ($\sim 25 - 90$ rad m$^{-2}$) higher than the galactic
contribution seems not to be related to the mini-halo\cite{brebru03}.

Other examples of mini-halos are in $PKS0745-191$\cite{bauode91},
Virgo\cite{owe00}, and possibly A2390\cite{bac03}.  The mini-halo in
A2390 is polarized at levels of $10-20$\%.

Gitti et al.\cite{git02} suggested that the electrons of the Perseus mini-halo
cannot be supplied by the central radio galaxy, but
are continuously undergoing reacceleration due to the MHD turbulence
associated with the cooling flow region.
They show that an isotropic magnetic field compression\cite{tri93} 
appears to well reproduce the observed surface brightness profile and 
total synchrotron spectrum along with the radial
spectral steepening. On the other hand the radial compression of
the magnetic field\cite{soksar90} does not appear to be applicable
to the mini-halo in the Perseus cluster. 
The above model was successfully applied also to the mini-halo in 
A2626\cite{git04}. Pfrommer and En{\ss}lin\cite{pfrens04}, on the other hand,
discussed the possibility that relativistic electrons in mini-halos
are of secondary origin and thus produced from the interaction of
cosmic ray protons with the ambient thermal protons.

\subsection{Magnetic fields beyond clusters}

Recent attempts to detect intergalactic magnetic fields beyond
clusters, i.e. in even more rarefied regions of the intergalactic
space, have shown recent promise in imaging diffuse synchrotron
radiation of very low level.  A very faint emission has been detected
at 327 MHz with the WSRT in the Coma
cluster region between the radio halo Coma C and 
the relic 1253+275\cite{kim89}. 
The surface brightness of this diffuse emission is very low and it is only
enhanced at low frequency and low resolution, so it is only 
visible as a positive noise in Fig. \ref{fig3}. The existence of this 
feature is confirmed by the  asymmetric extension
of the central halo Coma C imaged at 1.4 GHz with the Effelsberg
single dish\cite{dei97}, and by the VLA data at 74 MHz\cite{ens99a}.
The equipartition magnetic field in this region\cite{kro03} is 
$\simeq$ 10$^{-7}$G. These data may
indicate the existence of a more widespread and somewhat lower
intergalactic magnetic field than in the ICM.
It could be possibly associated with large-scale
shocks related to the formation of the large-scale structure in the
universe.

A possible evidence of magnetic field in the intergalactic medium
is found\cite{bag02} in the filament of galaxies ZwCl 2341.1+0000
at z $\sim$ 0.3.

The study of the RM of distant quasars could provide an independent
information on the magnetic field in the intergalactic medium.  An
upper limit of \ltsim 10$^{-9}$ G for a cosmic magnetic field outside
clusters of galaxies has been derived in the
literature\cite{kro94,val90}.  However, this limit relies on several
assumptions.  New generation instruments will shed light on this
point.

\section{Hard X-Ray Emission in Clusters}

\begin{figure}[th]
\centerline{\psfig{file=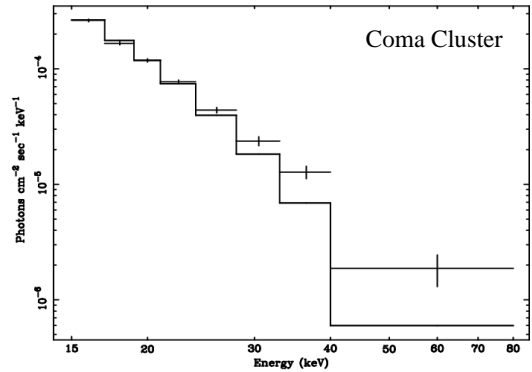,angle=0,width=7.5cm}}
\vspace{1cm}
\caption{
Coma cluster
--- PDS data. The continuous line represents the best fit with a
thermal component at the average cluster gas temperature of 8.1
keV. The errors bars are quoted at the
1$\sigma$ level$^{114}$.}
\label{fig5}
\end{figure}

Diffuse radio sources are not the only indication of non-thermal
activity in the ICM. The prospects for the X-ray detection 
of inverse-Compton emission originating from radio emitting electrons 
and photons of the microwave background were presented about 30 years 
ago\cite{hargri79,rep79}.  
Significant progress in the search of
non-thermal emission in the hard X-ray band ($>$ 20 keV, HXR) 
has been recently made 
owing to the improved sensitivity and wide spectral
capabilities of the {\it BeppoSAX} and the {\it Rossi X-ray Timing Explorer}
(RXTE) satellites (see the review by Fusco-Femiano et al.\cite{fus03a}).

Evidence for the presence of HXR radiation in excess to 
the thermal emission has been obtained in the spectrum 
of Coma\cite{fus99,rep99,repgru02,fus04}.
In Fig. \ref{fig5}, we report
the combined hard X--ray spectrum of the Coma cluster, obtained using
{\it BeppoSAX} data of two independent observations of 90 ksec and 300
ksec\cite{fus04}. 
The non-thermal excess with respect to the thermal emission is at a confidence
level of $\sim$ 4.8$\sigma$ and has a
flux of $(1.5\pm 0.5)\times 10^{-11}$ erg cm$^{-2}$ s$^{-1}$ in the
20--80 keV energy band (assuming a photon index $\Gamma_X$ = 2.0).  In
the framework of the IC model (see Sec. 3.4) the combination of the
radio and non-thermal X--ray fluxes allows an estimate of the
volume-averaged intracluster magnetic field of $\sim$ 0.2 $\mu$G.

In addition to Coma, HXR non-thermal emission has been detected in
A2256\cite{fus00,repgru03}.  The flux in the 20 -- 80 keV energy range
is $\sim 9\times 10^{-12}$ erg cm$^{-2}$ s$^{-1}$. A magnetic field of
$\sim$ 0.05 $\mu$G is derived for the northern cluster region, where
the radio relic is located, while a higher field value, $\sim$ 0.5
$\mu$G, could be present at the cluster center, in the region of the
radio halo.  A HXR detection at low confidence level is obtained in
A754\cite{val99,fus03b}, where however a radio galaxy 
with BL Lac characteristics could be responsible for the emission.

The detection in A2199, which is a cooling core cluster with no
extended diffuse radio emission, is controversial\cite{fus03a,kaa99}.
A marginal detection has been obtained in A119\cite{fus03a}, a merging
cluster without a radio halo, but the presence of several point
sources in the field of view makes the IC interpretation unlikely.

For the clusters A3667\cite{fus01} and A2163\cite{fer01} only
upper limits to the non-thermal X-ray emission have been derived.  
A possible detection in A2319 with RXTE\cite{grurep02}
leads to a magnetic field of $0.1-0.3$ $\mu$G.
Localized IC emission associated with the
radio relic and with merger shocks in A85 has been claimed from 
ROSAT observations\cite{bag98}. The derived 
magnetic field is $\sim$ 1 $\mu$G. 

The value of the magnetic field derived in the Coma
cluster by the IC HRX emission 
is quite consistent with that obtained
by the radio halo Coma C under equipartition conditions
(Sec. 4.1), but 
it is much lower than that derived from RM measurements (see next section). 
Therefore, alternative interpretations to the IC model for the 
non-thermal radiation detected in the Coma cluster have been proposed.
A suggested mechanism is the non-thermal bremsstrahlung
from supra-thermal electrons formed through the current acceleration 
of the thermal gas\cite{ens99b,dog00,sarkem00,bla00}. 
However, Petrosian\cite{pet01} pointed out that due to the low
efficiency of the bremsstrahlung mechanism, these models would require an 
unrealistically high energy input.

We will show in Sec. 9 that the disagreement between IC and RM
magnetic field measurements can be at least partially alleviated.
Future studies of non-thermal X-ray emission in clusters will 
be possible with the {\it ASTRO-E2} satellite.

\section{Rotation Measures}

One of the key techniques used to obtain information about the cluster
magnetic fields strength and geometry is the Faraday rotation analysis
of radio sources in the background of clusters or in the galaxy
clusters themselves.

RM calculated from extragalactic radio sources can be considered as
the sum of three integrals which represent the contribution of three
different regions, namely internal to the source itself, due to our
own Galaxy, and occurring in the ICM. The latter is 
the RM in which we are interested here.

Typical values of the RM of Galactic origin are of the order of 10 rad
m$^{-2}$ for most sources, and up to $\simeq 300$ rad m$^{-2}$ for
sources at low Galactic latitudes\cite{sim81}.  Once the contribution
of our Galaxy is subtracted, however, the RM of radio galaxies located
inside or behind clusters should be dominated by the contribution 
of the ICM.
   
High resolution RM studies of Cygnus A\cite{dre87} were the first to
demonstrate that the high RM values with large gradients on arcsec
scales cannot be either of Galactic origin or due to a thermal gas
mixed with the radio plasma, but must arise in an external screen of
magnetized, ionized plasma. Similarly, the asymmetric depolarization
found in double radio lobes embedded in galaxy clusters can be
understood as resulting from a difference in the Faraday depth of the
two lobes\cite{lai88,gar88,garcon91}(Laing-Garrington effect).
Indeed, the radio source lobe pointing towards the observer is less
depolarized than the lobe pointing away.

The observing strategy to get information on the cluster magnetic
field intensity and structure is to obtain high resolution RM maps of
sources located at different impact parameters of a cluster, then
derive the average value of the rotation measure \rmm~ and the
value of its dispersion \srm.  As described in Sec. 3.5.2, the RM
values are combined with measurements of the thermal gas density $n_e$
to estimate the cluster magnetic field along the line of sight.  Such
studies have been carried out on both statistical samples and on
individual objects.

The first successful statistical demonstrations of Faraday rotation
from radio sources seen through a cluster atmosphere were presented by
Lawler and Dennison\cite{lawden82} for a dozen of radio galaxies
and by Vall\'ee et al.\cite{val86}
for A2319. In both studies, a broadening of the values of RM was found
in the cluster sources, with respect to the sources in a control
sample.

Kim et al.\cite{kim90} investigated the magnetic field in the Coma
cluster using 18 radio sources, and found a significant enhancement of
the RM in the inner parts of the clusters.  They deduced a field
strength of $\sim$ 2 $\mu$G. 
For the magnetic field structure, they assumed the simple model with 
a single typical length for field reversal, i.e. a
cluster field consisting of cells of uniform size, with 
the same electron density and magnetic field strength, but with a 
random field orientation.
They obtained a cell size in the range 10 $-$ 30 kpc. In the following
year, Kim et al.\cite{kim91} improved the statistics by analyzing a
much larger sample of 106 radio sources, and deduced that magnetic
fields strengths in the cluster gas are of the order of 1 $\mu$G.  In
a more recent statistical study, Clarke et al.\cite{cla01} analyzed
the RMs for a representative sample of 27 cluster sources, plus a
control sample, and found a statistically significant broadening of
the RM distribution in the cluster sample, and a clear increase in the
width of the RM distribution toward smaller impact parameters.  Their
estimates give a magnetic field of $4 - 8$ $\mu$G, assuming a cell
size of $\sim$ 15 kpc.

The first detailed studies of RM on individual clusters have been
performed in cooling core clusters, owing to the presence of powerful
radio galaxies at their centers.  Extreme values of RMs are found to
be associated with these radiogalaxies, with the magnitude of the RMs
roughly proportional to the cooling rate.\cite{tay02}. Magnetic
fields, from $\sim$ 5 $\mu$G up to the values of $\sim$ 30 $\mu$G are
deduced in the innermost regions of these clusters, 
e.g. Hydra A\cite{tayper93} and 3C295\cite{pertay91,all01}.

\begin{figure}
\centerline{\psfig{file=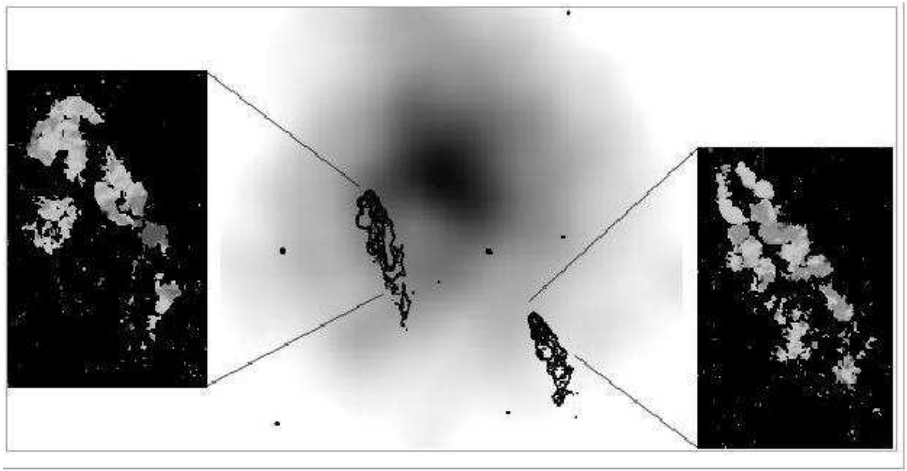,width=12cm}}
\caption
{VLA contour plot at 21 cm and RM images (insets) 
of the two tailed radio
galaxies 0053-015 (left) and 0053-016 (right) in A119$^{140}$. 
The contour plot is overlaid 
onto the ROSAT X-ray image (gray-scale) of the cluster.
The two radio galaxies are located at 
a projected distance from the cluster center
of $\sim$ 0.45$r_c$ and 1.2$r_c$ respectively.
Both the RM images show fluctuations on small scales ($\sim$ 10 kpc).
The RM values in 0053-015 are between $-$350 rad m$^{-2}$ and 
+450 rad m$^{-2}$ 
with \rmm = 28 rad m$^{-2}$, and \srm = 152 rad m$^{-2}$.
The RM values in 0053-016 are between $-$300 rad m$^{-2}$ 
and +200 rad m$^{-2}$,
with \rmm = --79 rad m$^{-2}$ and \srm = 91 rad m$^{-2}$.}
\label{fig6}
\end{figure}

Polarization data from sources at different cluster locations have been
obtained in clusters without cooling cores, i.e.  Coma\cite{fer95},
A119\cite{fer99b}, A514\cite{gov01c},
3C129\cite{tay01}, A400\cite{eilowe02}, A2634\cite{eilowe02}. 
In the Coma cluster, Feretti et al.\cite{fer95} 
derived a magnetic field of 7 $\mu G$ tangled on
scales of $\sim$ 1 kpc, in addition to a weaker field component of
$\sim$ $0.2$ $\mu$G, ordered on a scale of about one cluster core
radius. Generally, a decreasing
\absrmm~ and \srm~ with an increasing projected distance from the
cluster center is found. RM gradients are detected across the sources, 
indicating the presence of structure in the intracluster magnetic field.
The data lead to magnetic field estimates of $\sim$ $2-8$ $\mu$G, with patchy
structures of $\sim$ $5-15$ kpc. 

Overall, the data are consistent with cluster atmospheres containing
$\mu$G fields, with perhaps an order of magnitude scatter in field
strength between clusters, or within a given cluster, and with extreme
field values in cluster cooling cores. 
These estimates of the magnetic field strength from RM data crucially
depend on the magnetic field structure and geometry. The RM
distribution is generally patchy, indicating that large-scale magnetic
fields are not regularly ordered on cluster scales, but have structures on
scales as low as 10 kpc or less.
In Fig.\ref{fig6} we show the RM images obtained for the two 
central radio galaxies in A119\cite{fer99b}.

In many cases, high resolution RM images show a nearly Gaussian RM
distribution, suggesting an isotropic distribution of the field
component along the line-of-sight. However, many RM distributions show
clear evidence for a non-zero mean \rmm~ if averaged over areas
comparable with the radio source size, even after the Galactic
contribution is subtracted.  These \rmm~ offsets are likely due to
fluctuations of the cluster magnetic fields on scales greater than the
typical source size, i.e. considerably larger than those responsible
for the RM dispersion.  The random magnetic field must therefore both
be tangled on sufficiently small scales, in order to produce the
smallest structures observed in the RM images and also fluctuate on
scales one, or even two, orders of magnitude larger, to account for
the non-zero RM average.  For this reason, it is necessary to consider
cluster magnetic field models where both small and large scale
structures coexist.

So far very little attention has been given in the literature to the
determination of the power spectrum of the intracluster magnetic field
fluctuations.  Very recently En{\ss}lin and Vogt\cite{ensvog03} and
Vogt and En{\ss}lin\cite{vogens03} pointed out that the single scale
cell model is not realistic because it does not satisfy the condition
{\bf div}$\vec{B}$ = 0.
By using a semi-analytic technique, they
showed that the magnetic field power spectrum can be estimated by
Fourier transforming RM maps if very detailed RM images are available.
Moreover, they derived that the autocorrelation length of the RM
fluctuations is in general larger than the magnetic field
autocorrelation length.

An alternative numerical approach to investigate the strength  
and structure of cluster magnetic fields through Monte Carlo
simulations is presented in Murgia et al.\cite{mur04}.
A brief description of the capability of such a numerical 
approach is presented in Sec. 10.

It is worth mentioning here that some authors have suggested the
possibility that the RM observed in radio galaxies is not
associated with the foreground ICM, but may arise locally to the radio
source\cite{bic90,rudblu03}, 
either in a thin skin of dense warm gas mixed along the edge of
the radio emitting plasma, or in its immediate 
surroundings.  There are, however, several
arguments against this interpretation:\\
i) the trends of RM versus the cluster impact parameter 
in both statistical studies and
individual cluster investigations,\\
ii) the relation between the RM and the cooling flow rate in relaxed
clusters\cite{tay02},\\
iii) the Laing-Garrington effect\cite{lai88,gar88,garcon91},\\
iv) statistical tests on the scatter plot of RM versus 
polarization angle, for the radio galaxy  PKS1246-410 \cite{ens03},\\
v) the very
consistent scenario drawn by all the results presented in this
section.\\
Thus, we conclude that local effects might give some
contribution to the RM, however the major factor
responsible for the Faraday
Rotation should be the ICM. Future high resolution RM studies with
the next generation radio telescopes (e.g. EVLA, LOFAR, SKA) 
should help in
distinguishing the local effects, as well as possible effects arising
internally to the radio sources.

\section{Cluster Cold Fronts}

The high sensitivity and resolution of the {\it Chandra} satellite has
allowed the detection in the clusters A2142 and A3667 of sharp
discontinuities in the X-ray surface brightness, the so called cold
fronts\cite{mar00,vik01a,vik01b}.  Ettori and Fabian\cite{ettfab00}
pointed out that the observed temperature jumps in A2142 require that
thermal conduction across cold fronts must be suppressed by a factor
of 100 or more, compared to the classical Spitzer value\cite{spi62}.
Actually, a tangled magnetic field has been found\cite{chacow98} to reduce
the thermal conductivity from the Spitzer value by a factor of 
order 10$^2$ - 10$^3$.

\begin{figure}
\centerline{\psfig{file=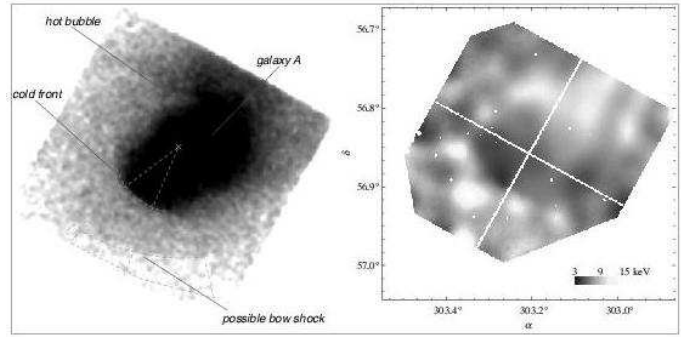,width=9cm}}
\caption
{Chandra X-ray image (left) and temperature map (right) 
of A3667$^{152,153}$.}
\label{fig7}
\end{figure}

Similar features have now been detected in several other
clusters\cite{mar03a,sun02,kem02}.  These structures are
apparently contact discontinuities between the gas which was in the cool
core of one of the merging sub-clusters and the surrounding intracluster
gas. They are not shocks because the density increase across the front is
accompanied by a temperature decrease such that there is no dramatic
change in the pressure and entropy.

In the cluster A3667\cite{vik01a,vik01b} (see Fig. \ref{fig7}),
the temperature discontinuity in the cold front 
occurs over a scale of 3.5\arcsec ($\simeq$ 4 kpc). 
Vikhlinin et al.\cite{vik01b} suggested that in order to reproduce 
such a sharp feature, magnetic fields are required to suppress
both thermal conduction and Kelvin-Helmholtz instabilities
along the contact discontinuity. 
They found that the front sharpness and its gradual smearing at large 
angles are most likely explained by the existence of a layer 
with a $\simeq$ 10 $\mu$G magnetic field parallel to the front.
The magnetic filed in the layer is probably
amplified by the stretching of the field lines.

\section{Magnetic Field Profile}

The simplest model for a cluster magnetic field is a  uniform field
through the whole cluster. However, this is not realistic: if the
field values detected at the cluster centers would extend over several
core radii, up to distances of the order of $\sim$ Mpc, the magnetic
pressure would exceed the thermal pressure in the outer parts of the
clusters.

\begin{figure}[th]
\centerline{\psfig{file=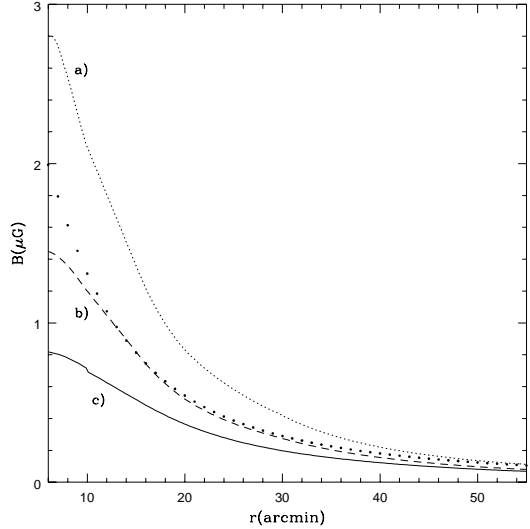,width=7cm}}
\caption
{The magnetic field profile in the Coma cluster obtained from the
radial spectral steepening, by applying an electron reacceleration
model. The figure is from Brunetti et al.$^{50}$, who use H$_0$ = 50
km s$^{-1}$ Mpc$^{-1}$.  The trends, however, do not depend on the
assumed cosmology.  Different lines refer to different values of the
reacceleration coefficient.  The dots represents the Jaffe$^{159}$
theoretical predictions for the magnetic field distribution in the
Coma cluster.}
\label{fig8}
\end{figure}

Jaffe\cite{jaf80} first suggested that the magnetic field distribution in a
cluster would depend on the thermal gas density and on the
distribution of massive galaxies and therefore would decline with the
cluster radius. Constraints to the radial gradient of the cluster
magnetic field strength are provided by observations of clusters
hosting a radio halo.  Indeed, the spatial correlation found in some
clusters between the X-ray cluster brightness and the radio halo
brightness\cite{gov01b}  
implies that the energy densities in the thermal and non-thermal
components have a similar radial scaling, thus a magnetic field
decline is inferred.

A radial decrease of the cluster magnetic field strength is also
deduced in the framework of halo formation models which consider the
reacceleration of the radio emitting electrons.  The radial steepening
of the synchrotron spectrum, observed in Coma\cite{gio93} 
and more recently in A665\cite{fer04} and A2163\cite{fer04}, is
interpreted as the result of the combination of the magnetic field
profile with the spatial distribution of the reacceleration
efficiency, thus allowing us to set constraints on the radial profile
of the cluster magnetic field.  
In Fig. \ref{fig8} we show the magnetic field profile in the Coma
cluster obtained by Brunetti et al.\cite{bru01} by applying a model
for the electron reacceleration. Different lines refer to different values of
the reacceleration coefficient. The dots represents the Jaffe\cite{jaf80}
theoretical prediction. The field intensity is found to
decrease smoothly from $\sim~0.5-1.5~\mu$G at the cluster center to 
$\sim~0.03-0.5~ \mu$G at $\sim$ 1.3 Mpc radius, 
with a trend similar to that of the thermal gas.
The magnetic field profiles in A665\cite{fer04} and A2163\cite{fer04}
show a flatter decline, probably because
the ongoing violent mergers in these clusters are playing a
significant role in
determining the conditions of the radiating particles and of the
magnetic field.

Important clues on the magnetic field distribution are also derived
from MHD cosmological simulations.  Dolag et al.\cite{dol99,dol02}
simulated the formation of magnetized galaxy clusters from an initial
density perturbation field, using a cosmological MHD code.  They found
that the $\mu$G level field presently observed in clusters can be
reproduced by the evolution of the magnetic field starting from an
initial field of $\sim 10^{-9}$ G at redshift 15. This field is
amplified by compression during the cluster collapse.  They obtained that
the process of large-scale structure formation in the universe drives
the characteristics of these magnetic fields.  One of their results is
that the magnetic field strength at any point within galaxy clusters
is proportional to the gas density.

In the simple case of adiabatic compression during a spherical
collapse due to gravity, the field lines are frozen into the plasma,
and compression of the plasma results in compression of flux lines.
The expected growth of the magnetic field is roughly proportional to
the gas density as B $\propto \rho^{2/3}$, as a consequence of magnetic flux
conservation.

From the simulations, Dolag et al.\cite{dol01} predict the existence
of a correlation between  the Faraday rotation measure and the X-ray 
flux.  They find that $\sigma_{RM}$ increases with the
X-ray flux:
\begin{equation}
\sigma_{RM}\propto S_{x}^f,
\end{equation}
\noindent
with $f \simeq 1$.

The X-ray surface brightness  is:
\begin{equation}
   S_{x} \propto
   \int n_{e}^2 \; \sqrt{T} \; dl.
\end{equation}
\noindent
The RM dispersion, obtained from Eq. \ref{sigmarndwalk},
is related to $B$ and $n_e$. 
The two observables $\sigma_{RM}$ and S$_{x}$ 
relate the two line of sight integrals
with each other, therefore in comparing these two quantities, 
we actually compare
cluster magnetic field versus thermal density. 
Thus the magnetic field profile can be represented
by:

\begin{equation}
B(r)\propto n_e(r)^\eta.
\label{kingB}
\end{equation}
\noindent
In the case of the $\beta$-model (Eq. \ref{king}), 
the X-ray flux $S_{x}$ is: 
\begin{equation}
S_{x} \propto (1+r_{\perp}^{2}/r_{c}^{2})^{-3\beta+{\bf \frac{1}{2}}}.
\end{equation}
\noindent
By substituting Eq.~\ref{kingB} in the expression of $\sigma_{RM}$ derived 
from Eq.~\ref{sigmarndwalk}, we obtain:
\begin{equation}
\sigma_{RM} \propto (1+r_{\perp}^{2}/r_{c}^{2})^{\bf {-\frac{3}{2}
\beta(1+\eta)+\frac{1}{4}}}
\end{equation}
\noindent
Thus, by comparing $S_{x}$ and 
$\sigma_{RM}$, one finds that the index $\eta$ is related to the 
slope $f$ and to the parameter $\beta$, 
through:
\begin{equation}
   \eta ={{1}\over{\beta}}(2f-1)(\beta-{{1}\over{6}})
\label{alpha}
\end{equation}
\noindent
We note that for a constant magnetic field ($\eta=0$)  
the slope of the $\sigma_{RM}-S_{x}$ correlation
should be $f = 0.5$ while a steeper slope would imply $\eta > 0$.  

For the cluster A119, where the polarization properties of
3 extended radio galaxies are available, the
$\sigma_{RM}$ -- S$_{x}$ relation  yields $B \propto
n_e^{\eta}$ with $\eta=0.9$.
This implies that the central magnetic field\cite{dol01} in this cluster
is $\approx$ 9 $\mu$G, instead of the $\approx$ 6 $\mu$G
inferred by using the constant magnetic field
approximation\cite{fer99b}.

\begin{figure}[th]
\centerline{\psfig{file=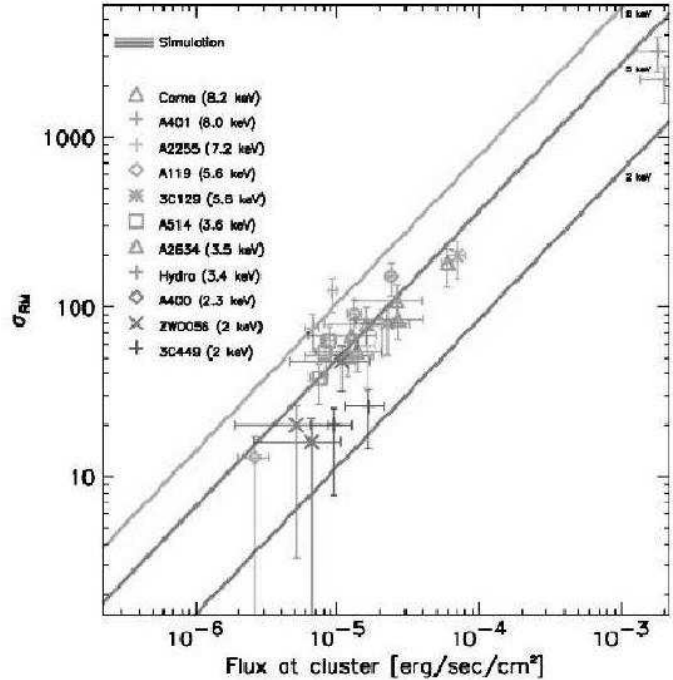,width=9cm}}
\caption
{
The correlation between the observed $\sigma_{RM}$ 
and the X-Ray flux, and comparison with
 theoretical predictions (kindly supplied by K. Dolag). 
 The data points are marked with different grey scale levels
 according to the individual cluster temperature. Theoretical
 predictions are shown as lines for three temperatures. Data points
 for different sources in the same cluster follow very well the
 predicted lines. The simulations also predict these lines to be
 shifted according to the temperature of the cluster. This trend is
 also confirmed by the data points.
}
\label{fig9}
\end{figure}

The simulations not only predict that the magnetic field scales
similarly to the density within all clusters but also that
 clusters should have different central magnetic 
field strengths depending on
their temperature, and therefore their mass.
Indeed, the normalization of the $\sigma_{RM}$ -- S$_{x}$ relation 
is expected to be related to the cluster temperature.
In Fig. \ref{fig9}, 
the predicted $\sigma_{RM}$ -- S$_{x}$ relation 
at different cluster temperatures 
has been compared with the observations 
for a sample of clusters with good RM data.

\section{Reconciling Magnetic Field Values}

\begin{table}[ht]
\tbl{Magnetic field estimates derived from various methods 
in the clusters Coma and  A3667.} 
{\begin{tabular}{@{}ccccc@{}} \toprule
  Name  & Method     & Field strength & Location      & Reference \\
        &            & ($\mu$G)       &               & \\ \colrule
 Coma   & Equipartition    & 0.45    & radio halo  &  34 \\
        & Equipartition    & 0.55    & radio relic &  77 \\
        & Faraday Rotation &  7   & cluster center &  16 \\
        & Faraday Rotation & 0.2 & cluster center(large scale) & 16  \\
        & Inverse Compton  & 0.2 & cluster average& 114 \\ \toprule
A3667   & Equipartition    & 1.5$-$2.5    & NW relic& 86 \\
        & Inverse Compton  & $\geq$ 0.4  & cluster average& 120  \\
        & Faraday Rotation & 1$-$2         & cluster center& 86  \\
        &  Faraday Rotation & 3$-$5       & NW relic&  86 \\
        & Cold front &  10         & along the cold fronts& 153 \\ \botrule
 \end{tabular}}
Column 2 gives the method used to estimate the field strength,
Column 3 the value of the magnetic field in $\mu$G, Column 4 describes the
location in the cluster at which this estimation is made,
Column 5 gives the reference.
\end{table}

From the results presented in the previous sections, it is derived 
that cluster magnetic field strengths obtained from RM arguments
(Sec. 6) are about an order of magnitude higher than the estimates
obtained from both the diffuse synchrotron
 radio halo emission under equipartition conditions (Sec. 4.1)
and the inverse Compton hard X-ray emission (Sec. 5). 
Relatively high magnetic fields could be present in regions of radio
relics at the cluster periphery (Sec. 4.2), and in cold fronts
(Sec. 7). However, it is important to note that magnetic field
estimates derived in relics and in cold fronts may be not
representative of the overall cluster magnetic field strengths because
they have been likely enhanced by compression.

The clusters Coma and 
A3667\cite{joh03} are unique in that they
allow the field to be estimated by using different techniques.
Examples of the variation of magnetic field strength estimates from
various methods and in various locations of these clusters are given
in Table 3.

Several arguments can be invoked to alleviate the discrepancies
between different methods of analysis. First, we remind that
the equipartition values rely on several assumptions (Sec. 3.2).  
Moreover, the radio synchrotron
and IC emissions originate from large cluster volumes, and the
corresponding magnetic field estimates are averaged over the whole
cluster, whereas the RM gives an average of the field along the line of
sight, weighted by the thermal gas distribution.  Taking into account
the radial profile of the cluster magnetic field and of the gas
density, Goldshmidt and Rephaeli\cite{golrep93} first showed that the
field strength estimated with the IC method is expected to be smaller
than that measured with the RM observations. 
Beck et al.\cite{bec03} pointed out that field estimates derived from RM
may be too large in the case of a turbulent medium
where small-scale 
fluctuations in the magnetic field and the electron density are highly
correlated.
Finally, more realistic
electron spectra should be considered in the analysis of synchrotron
and IC emission.  It has been shown that IC models which include both
the expected radial profile of the magnetic field, and anisotropies in
the pitch angle distribution for the electrons allow higher values of
the ICM magnetic field in better agreement with the Faraday rotation
measurements\cite{bru03,pet01}.  Moreover, as shown in Table 3,
the magnetic field strength may vary depending on the dynamical history
and the location within the cluster.

In some cases a radio source could compress the gas and fields in the
ICM to produce local RM enhancements\cite{bic90,rudblu03} (see also
Sec. 6), thus leading to overestimates of the derived ICM magnetic
field strength.
 
The magnetic field may show complex structure, as filaments and/or
substructure with a range of coherence scales, therefore the
interpretation of RM data as given in Sec. 3.5.2 would be too simplified.
Indeed, Newman et al.\cite{new02} demonstrated that the assumption of 
a single-scale magnetic field leads to an overestimation 
of the magnetic field strength calculated through
RM studies.
In the next section we show that the use of a numerical approach can
significantly improve our interpretation of the data and thus the
knowledge of the strength and structure of magnetic fields.

\section{Cluster Magnetic Fields Through a Numerical Approach}

We have seen that cluster magnetic field strengths can be calculated
through their effects on the polarization properties of radio galaxies
by using an analytical formulation (Sec. 3.5.2) based on the
approximation that the magnetic field is tangled on a single scale.
However, detailed observations of radio sources and MHD
simulations\cite{ensvog03,vogens03,dol02} 
suggest that it is necessary to consider more
realistic cluster magnetic fields which fluctuate over a wide range of
spatial scales. To accomplish this, Murgia et al.\cite{mur04}
simulated random three-dimensional magnetic fields with a power-law
power spectrum: $|B_{\kappa}|^2 \propto {\kappa}^{-n}$, where
${\kappa}$ represents the wave number of the fluctuation scale.  They
investigated the effects of the expected Faraday rotation on the
polarization properties of radio galaxies and radio halos, by
analyzing the rotation measure effects produced
by a magnetic field with a
power spectrum which extends over a large range of spatial scales
($6-770$ kpc) and with different values of the spectral index
$(n=2,3,4)$.

\subsection{Simulated rotation measures}

\begin{figure}
\centerline{\psfig{file=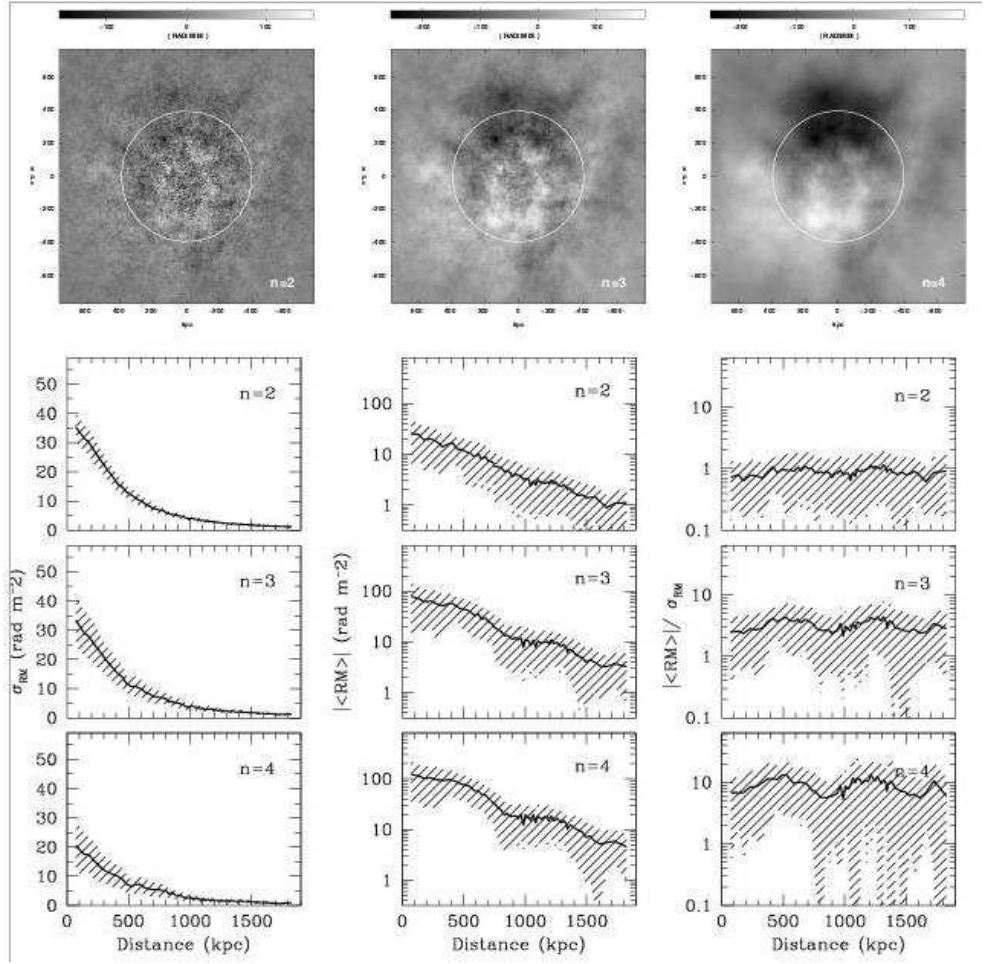,width=13cm}}
\caption{
Top: simulated RM images for magnetic 
field power spectrum spectral index $n=2,3,4$. 
The electron gas density of the cluster follow a standard
$\beta$-model with a core radius $r_{c}$ = 400 kpc 
(indicated by a circle in the figure) a central density 
$n_{e}(0)$=10$^{-3}$ cm$^{-3}$ and $\beta$ = 0.6.
The three power spectra are normalized to have the 
same total magnetic field energy which is distributed over the range 
of spatial scales from 6 kpc up to 770 kpc. 
The field at the cluster center is $B_{0}$ = 1 ~$\mu$G and its  
energy density decreases from the cluster center according 
to $B(r)^{2}\propto n_{e}(r)$.
Bottom: radial profiles (\srm, \absrmm~ and \absrmm/\srm~ respectively) 
obtained from the RM simulations described above.
The profiles have been 
obtained by averaging the simulated RM images in regions 
of 50 $\times$ 50 kpc$^2$, which is a typical size for radio galaxies.
The figure is from Murgia et al.$^{146}$} 
\label{fig10}
\end{figure}

Fig. \ref{fig10} (top) shows the 
simulated RM images obtained with different values
of the index $n$ for a typical cluster of galaxies (see caption for
more details).
Different power spectrum indexes will generate
different magnetic field configurations and therefore will give rise
to very different simulated RM images.  Fig. \ref{fig10} (bottom)
shows the simulated profiles of \srm, \absrmm, and
\absrmm/\srm~ (left, central and right panels, respectively), 
as a function of the projected distance from the
cluster center.  While both \srm~ and \absrmm~ increase linearly with
the cluster magnetic field strength, the ratio \absrmm/\srm~ depends
only on the magnetic field power spectrum slope, for a given range of
fluctuation scales.  Therefore the comparison between RM data of
radio galaxies embedded in a cluster of galaxies and the simulated
profiles, allows the inference of
both the strength and the power spectrum slope of
the cluster magnetic field.

Typical measured values of
\absrmm/\srm~ for cluster radio galaxies are derived to be in the
range 0.2 - 1.2, once the Galaxy contribution is
subtracted\cite{mur04}. 
Thus the comparison of the observations with the simulations 
leads to a rather flat cluster magnetic field power
spectrum, with a spectral index $n \simeq 2$. This indicates that most
of the magnetic energy density is on the smaller scales.
   
Another result of the simulations is that, when a power spectrum 
of the magnetic field
is assumed, the inferred  magnetic field strength 
is about a factor of 2 lower than the value computed 
from Eq.~\ref{felten} if the single scale
$\Lambda_{c}$ is taken to be equal to the smallest
patchy structures detectable in the RM images, as frequently used.
This implies that the magnetic fields derived from RM measurements
may be  overestimated (see Sec. 9).

\subsection{Simulated radio halo polarization}

Different values of the power spectrum spectral index will generate
very different total intensity and polarization brightness
distributions for the radio emission of a halo.
So far, polarization emission from radio halos has not been
detected. The current upper limits to the polarization at 1.4 GHz are
a few percent ($\sim 5$\%).

Murgia et al.\cite{mur04} simulated  the expected halo total intensity 
and polarization brightness distributions
at 1.4 GHz and 327 MHz, as they would appear
when observed with a beam of 45\arcsec, 
by introducing in
the 3-dimensional magnetic field an isotropic population of
relativistic electrons.
Different values of the magnetic field strength and power spectrum index
were assumed.

\begin{figure}
\centerline{\psfig{file=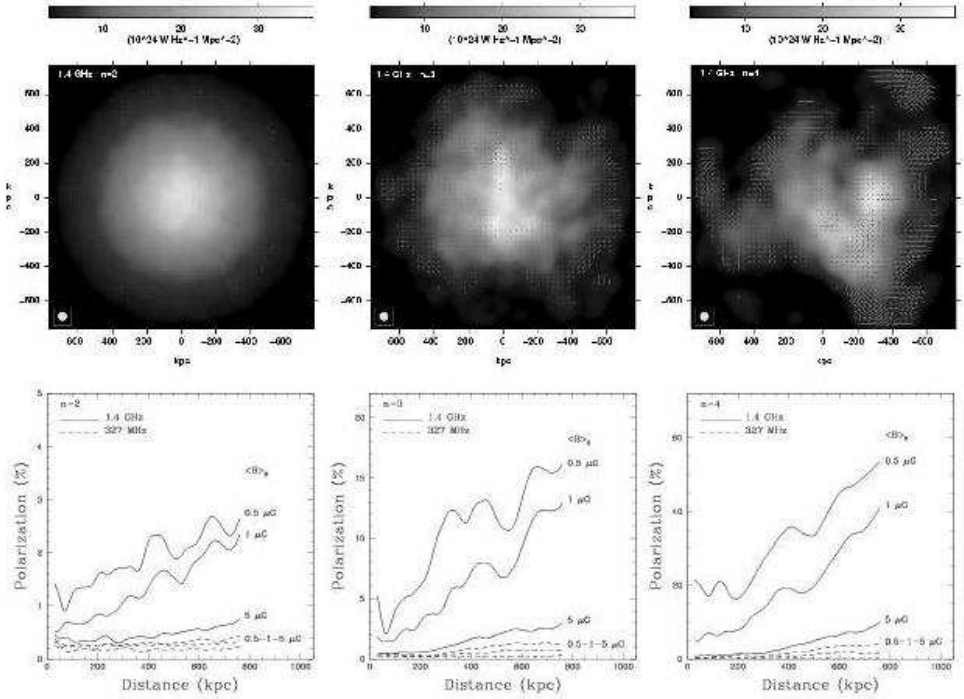,width=13cm}}
\caption
{Simulated halo brightness and polarization degree$^{146}$
for cluster at
$z=0.075$ as it would appear when observed with a beam of 45\arcsec.  
Top: simulated halo images at 1.4
GHz for different values of the magnetic field power spectrum slope
$n$ and $B_{0}$=1 $\mu$G; the vectors lengths are
proportional to the degree of polarization, with 100 percent
corresponding to 100 kpc on the sky.  Field directions are those of
the E-vector.  Bottom: radially averaged profiles of the polarization
percentage at 327 MHz and 1.4 GHz for three values of the magnetic
field strength, namely $\langle B\rangle_{0}= $ 0.5, 1 and 5 $\mu$ G.}  
\label{fig11}
\end{figure}

Fig. \ref{fig11} (top) shows simulated radio halo brightness and 
polarization percentage distributions at 1.4 GHz 
(see caption for more details).
Fig. \ref{fig11} (bottom) shows the expected fractional 
polarization profiles at 1.4 GHz and 327 MHz for the different values of the average magnetic field strength and power spectrum spectral index. 
Simulations indicate that a power spectrum slope steeper than $n=3$ and
a magnetic field strength lower than $\sim 1 \mu$G
result in a radio halo polarization percentage at  1.4 GHz
far in excess of the current observational upper limits.
This means that, in agreement with the RM simulations, 
either the power spectrum spectral index is flatter
than $n=3$ or the magnetic field strength is significantly higher than
$\sim 1 \mu$G. The halo depolarization at 327 MHz is 
particularly severe and the expected polarization percentage
at this frequency is always below 1\%. 
Moreover it is also evident that the magnetic field power
spectrum slope has a significant effect in shaping the radio halo. In
particular, flat power spectrum indexes ($n<3$) give raise to
very smooth radio brightness images (under the assumption that the
radiating electrons are uniformly distributed). 

\section{Origin and Amplification of Cluster Magnetic Fields}

The origin of the magnetic fields observed in galaxies and clusters of
galaxies is debated.  
Very little is known about their existence before and after
the time of recombination, their evolution, and the possible impact
they could have on galaxy and structure formation.
We very briefly give the outlines of the main scenarios proposed for
the magnetic fields in the ICM, without going into
the details, which can be found in the 
literature\cite{cartay02,wid02}. 

According to the first scenario, 
cluster magnetic fields may be {\it primordial}, i.e. generated in
the early universe prior to recombination\cite{grarub01}.  
In this case, magnetic
fields would be already present at the onset of structure formation,
and would be a remnant of the early Universe.  One mechanism for the
generation of primordial fields involves the ``Biermann battery''
effect\cite{bie50}, which occurs when the gradients of
electron pressure and number density are not parallel, thus
electrostatic equilibrium is no longer possible. This leads to a
thermoelectric current which generates an electric field (and a
corresponding magnetic field) that restores force balance. Other
possibilities might be that weak seed fields were formed in the phase
transitions of the early Universe, such as a quark-hadron (QCD), or
electroweak (EW) transition, where local charge separation occurs
creating local currents, or during inflation, where electromagnetic
quantum fluctuations are amplified\cite{sto02}.  
Values of these seed fields are of the order of $\sim$
10$^{-21}$ G.

In principle, the presence of magnetic fields in the very early
Universe might be detectable through their effect on the Big Bang
nucleosynthesis, or if the expansion is observed to be anisotropic.
Current observations of anisotropy in the CMB place weak upper limits
(B $<$ 5 $\times$ 10$^{-9}$ G) on the
strength of a homogeneous component of a primordial magnetic field
generated in this way\cite{bar97}.  By analyzing the
effect of the inhomogeneities in the matter distribution of the
universe on the Faraday rotation of distant QSOs, 
limits of B $<$ $10^{-9}-10^{-8}$ G are obtained\cite{bla99}, 
depending on the assumed scales of the fluctuations.

Another scenario is that the cosmological magnetic fields are
generated in later epochs of the Universe. Gnedin et al.\cite{gne00}
argued that the strongest ``Biermann battery'' effects are
likely to be associated with the epoch of cosmological
reionization. Kulsrud et al.\cite{kul97} investigated the
possibility that the  field may be {\it protogalactic}, i.e. 
 generated during the initial stages of the structure
formation process, during the protogalaxy formation.

A third scenario  involves
the {\it galactic} origin, i.e.  ejection from galactic winds of
normal galaxies or from active and starburst galaxies\cite{kro99,volato99}.
Galaxy outflows, gas stripping,
ejection from the AGN by radio jets, all contribute to deposit
magnetic fields into the ICM. Galactic fields may be arise from the
fields in the earliest stars, then ejected into the interstellar
medium by stellar outflows and supernova explosions.  Alternatively,
fields in galaxies may result directly from a primordial field that is
adiabatically compressed when the protogalactic cloud collapses.
Indeed, battery mechanism on galactic scales can generate fields up to
10$^{-19}$ G.

Support for a galactic injection in the ICM comes from the evidence
that a large fraction of the ICM is of galactic origin, since it
contains a significant concentration of metals.  However, fields in
clusters have strengths and coherence size comparable to, and in some
cases larger than, galactic fields\cite{grarub01}. 
Therefore, it seems quite difficult that the magnetic fields in
the ICM derive from ejection of the galactic fields.  The recent
observations of strong magnetic fields in galaxy clusters suggest that
the origin of these fields may indeed be primordial.

The observed field strengths greatly exceed the amplitude of the seed
fields, or of fields injected by some mechanism.  Therefore, magnetic
field amplification seems unavoidable.  Dynamo effect can be at
work. A magnetic dynamo consists of electrically conducting matter
moving in a magnetic field in such a way that the induced currents
maintain and amplify the original field\cite{bec96}.  The essential
features of the galactic dynamo model are turbulent motions in the
interstellar medium, driven by stellar winds, supernova explosions,
and hydromagnetic instabilities.  

In addition, amplification can occur during the cluster collapse.
During the hierarchical cluster formation process, mergers generate
shocks, bulk flows and turbulence within the ICM.  The first two of
these processes can result in some field amplification simply through
compression. However, it is the turbulence which is the most promising
source of non-linear amplification.  MHD calculations
have been performed\cite{dol99,sch02,roe99b} to
investigate the origin, distribution, strength and evolution of the
magnetic fields.  The results of these simulations show that cluster
mergers can dramatically alter the local strength and structure of
cluster-wide magnetic fields, with a strong amplification of the
magnetic field intensity.  The initial field
distribution at the beginning of the simulations at high redshift is
irrelevant for the final structure of the magnetic field. The final
structure is dominated only by the cluster collapse. Fields
can be amplified  from
values of $\sim$ 10$^{-9}$ G to $\sim$ 10$^{-6}$ G.

\begin{figure}[th]
\centerline{\psfig{file=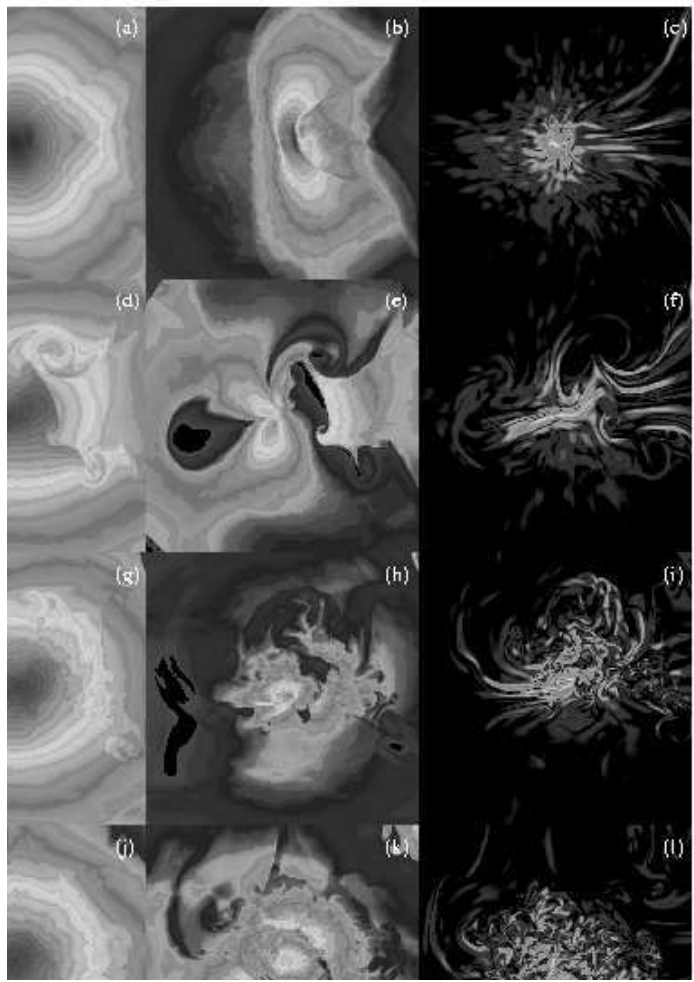,bb=77 77 535 690,clip,width=11cm}}
\caption
{
Three-dimensional numerical MHD simulations of
magnetic field evolution in merging clusters of galaxies$^{174}$.
The evolution of gas density (column 1),
gas temperature (column 2), and magnetic pressure (column 3) in
two-dimensional slices taken through the cluster core in the plane of
the merger. The four rows represent  different epoch during the
merger: $t$ = 0,1.3, 3.4,
and 5.0 Gyr, respectively. 
}
\label{fig12}
\end{figure}

Roettiger et al.\cite{roe99b} found a significant evolution
(see Fig. \ref{fig12}) of the structure
and strength of the magnetic fields during two distinct epochs of the
merger evolution. In the first, the field becomes quite filamentary as
a result of stretching and compression caused by shocks and bulk flows
during infall, but only minimal amplification occurs. In the second,
amplification of the field occurs more rapidly, particularly in
localized regions, as the bulk flow is replaced by turbulent motions.
Shear flows are extremely important for the amplification of the
magnetic field, while the compression of the gas is of minor
importance. Mergers change the local magnetic field strength
drastically. But also the structure of the cluster-wide field is
influenced. At early stages of the merger the filamentary structures
prevail. This structure breaks down later ($\sim$ 2--3 Gyr) and leaves
a stochastically ordered magnetic field.


\section{Conclusions}
Our knowledge of the magnetic field properties in galaxy clusters 
has significantly improved in recent years, owing to the improved
capabilities of radio and X-ray telescopes.
It is well established that $\mu$G level magnetic fields are
widespread in the ICM, regardless of the presence of large-scale
diffuse radio emission. The magnetic field strengths show
almost an order of magnitude scatter between clusters, or within a
given cluster, and are extreme in cluster cooling cores.
For such
large fields the magnetic pressure is comparable to or larger than the
gas pressure derived from X-ray data, suggesting that magnetic fields
may play a significant role in the cluster dynamics.

The observations are often interpreted in terms of the simplest
possible model, i.e. a constant field throughout the
whole cluster.  However, a decline with radius is expected if the
intensity of the magnetic field results from the compression of the
thermal plasma during the cluster gravitational collapse.  Observational
evidence of magnetic field profiles has been derived in some clusters.
Moreover the magnetic field could show complex structure with a 
range of coherence scales. 

The study of cluster magnetic fields has gained a big interest in
recent years, leading to several new observations as well as
simulations.  There are, however, still many questions to answer: are
the fields filamentary, what are the coherence scales, to what extent
do the thermal and non-thermal plasmas mix in cluster atmospheres, how
do the fields extend, what is the radial trend of the field strength,
how does the field strength depend on cluster parameters such as the gas
temperature, metallicity, mass, substructure and density profile, how
do the fields evolve with cosmic time, and finally how were the
fields generated?

New generation instruments in the radio band, as the EVLA, LOFAR and SKA,
 are rather promising and will establish a clear
connection between radio astronomical techniques and the improvement in
the knowledge of the X-ray sky. There are various satellite missions,
as {\it ASTRO-E2}, {\it XEUS} and {\it Constellation X}, 
which will map the X-ray sky at low and high 
 energies in the next years. These
will provide a more precise knowledge of the X-ray surface brightness
of clusters, i.e. of their thermal gas density, allowing
a more accurate and correct interpretation of  the sensitive RM 
measurements. The detection of HXR non-thermal
emission will provide independent measurements of the magnetic fields.
The accurate experimental determination of large-scale
magnetic fields in the intracluster medium will thus be possible.
The detection of synchrotron radiation at the lowest possible levels
will allow the measurement of magnetic fields in even more rarefied
regions of the intergalactic space, and the investigation of the
relation between the formation of magnetic fields and the formation
of the large-scale structure in the universe.

\section*{Acknowledgments}

We are grateful to Rainer Beck, Gianfranco Brunetti, Tracy Clarke, 
Klaus Dolag, Torsten En{\ss}lin, Roberto Fanti, Roberto Fusco-Femiano, 
Gabriele Giovannini, Dan Harris, Melanie Johnston-Hollitt, 
Phil Kronberg, Matteo Murgia, Greg Taylor,
and Corina Vogt for several fruitful discussions on this topic,
and for suggestions.
We are indebted to Klaus Dolag for supplying Fig. \ref{fig9}.

\end{document}